**The Capability Transfer: Cohort Reversal, Codified Expertise, and the Adaptability Tax of AI Transformation**

Ina Catrinescu

Future Positive, Amsterdam

**Abstract**

Caosun and Aral (2026) show that a decision-maker who chooses AI usage intensity for a worker can rationally trade long-run skill for short-run output, and that misaligned incentives turn this steady-state loss into an augmentation trap. We extend their model from one worker to a firm with two cohorts and a task distribution ordered by tacitness, and add one state variable, the codified frontier: the share of the firm's tasks on which AI supplies a skill-neutral productivity floor. The extension rests on one assumption, that the tacit capability a tool supplies is produced by unassisted practice with stakes, and that the deployment which supplies it displaces the practice that produced it. We state the assumption at two altitudes. At the level of the market it is a claim about the source population of a shared model, and the evidence for it is the collapse of the public commons that fed the first models; at the level of the firm it applies only to the firm-specific residual of the tacit set, which no vendor can codify and no vendor can regenerate. Four results follow. First, adoption compresses junior demand before senior demand, because the codified tasks were the junior's job; calibrated to the 2023-2025 productivity experiments, with the adoption curve imposed, the model places the junior trough in 2027 in every calibration in which junior headcount reaches a trough, and reproduces the direction of the early-career employment gap in payroll data through mid-2026. Second, once adoption saturates the frontier keeps moving, largely through the vendor, and the cohort composition reverses: the junior share of headcount rises from 58 to 66 percent by 2035 in the baseline, in which the firm captures nothing of its own seniors' judgment, and to 75 percent in a firm that does; whether required senior headcount ever halves is governed by a closed-form condition on the vendor's codification rate, the firm's capture rate and the task-renewal rate, and in the baseline it does not. Third, senior formation can occur only on tasks the tool does not cover, and that set shrinks with the frontier, so the career ladder inverts: a firm that does not deliberately assign uncovered work to juniors, and place answerability for that work with them, stops forming seniors of its own. The claim concerns how a firm allocates work and consequence, and not whether experts can be formed at all. Fourth, the reversal is cheap in output and expensive in capability, and the output cost arrives through the pipeline rather than the frontier. In the baseline the output loss is zero until the senior stock falls below requirement in 2032, 1.5 percent of task mass by 2035 and 9 percent by 2050, and none of it falls on the firm's own residual before 2050; what falls early is the firm's capacity on uncovered work, by 41 to 79 percent by 2035 to 2050, and its ability to rebuild its own frontier after a renewal shock. We price the transfer from three sides using a receiving-end build-cost model: the saving is immediate and visible, 93 percent of the capability the tool now substitutes was financed by the workers who built it, and we state the loss as a distributional fact rather than an efficiency claim.

This paper replicates and extends the model in Caosun, M. and Aral, S. (2026), *The Augmentation Trap: AI Productivity and the Cost of Cognitive Offloading*, arXiv:2604.03501v1, 3 April 2026. All quotations, section references and replication targets in this paper are to v1; that paper has since been revised, and readers using a later version should expect the section numbering and some wording to differ. All propositions, notation and closed forms attributed to that paper are theirs; the cohort extension, the calibration, the empirical section and the ledger are ours. Replication code, the simulation and the companion workbook are available with this paper.

## 1 Introduction

Unemployment among recent college graduates in the United States stood at about 5.6 percent in the second quarter of 2026, with underemployment at 42 percent (Federal Reserve Bank of New York, 2026). Payroll microdata show employment of 22 to 25 year olds in the occupations most exposed to generative AI running 19 percent below where it would be had it kept pace with less-exposed occupations, a shortfall that widened from 15 percent a year earlier, while experienced workers in the same occupations show no comparable gap (Brynjolfsson et al., 2026). Entry-level job postings on Indeed were 7.5 percent lower in May 2026 than a year earlier, senior-level postings 14.7 percent higher, and workers with more than ten years' experience were the largest single group applying to entry-level roles (Indeed Hiring Lab, 2026a,b). Euro area youth unemployment rose to 15.1 percent in the first quarter of 2026, with youth employment in ICT down 18.6 percent over three years (European Central Bank, 2026). The OECD reports that youth employment rates fell between 2024 and 2026 while workers aged 55 to 64 supplied two-thirds of employment growth (OECD, 2026).

The official interpretations of these facts are cautious. The ECB attributes euro area weakness to cyclical cooling, the OECD finds that the role of large language models in explaining young people's difficulties "appears to be limited", and the ILO counts 6.1 percent of youth employment in occupations highly susceptible to AI-related disruption (ILO, 2026). The payroll evidence is more specific: the decline is concentrated in occupations where AI automates rather than augments, it operates through hiring rather than wages, and it falls on workers whose knowledge is codified rather than tacit (Brynjolfsson et al., 2025, 2026). This paper takes that specificity seriously and asks what the same mechanism implies for the other end of the seniority distribution.

Our starting point is the dynamic model of Caosun and Aral (2026). In that model a decision-maker chooses AI usage intensity for a worker, trading immediate productivity against the erosion of the skill on which the productivity of AI partly depends. The model's central object is a decomposition of the tool's productivity effect into a skill-neutral component α and a knowledge-complementary component β, and its central result is that adoption can be privately rational and long-run harmful at once. The model holds α and β fixed and treats one worker in isolation. Its authors note both restrictions and, on the second, write that "if AI erodes a particular skill across many workers simultaneously, the scarcity value of that skill changes", and "training pipelines, apprenticeships, and residency programs depend on experienced practitioners who can teach; once they deskill, there are not enough left to train the next generation" (Caosun and Aral, 2026, Section 3.5).

We relax both restrictions in the smallest way that connects them. We let α vary over time, and we let it vary because of what workers do. The knowledge on which a language model operates is codified human output: text, code, procedures, corrected drafts, logged decisions. The payroll evidence is "consistent with a world in which generative AI is particularly effective at reproducing and applying knowledge that has already been encoded in text and other digital information" (Brynjolfsson et al., 2026). The public commons that supplied the first generation of that knowledge has stopped being replenished: monthly new questions on Stack Overflow, above 200,000 at the 2014 peak and still in the tens of thousands when the first mass-market model was released, fell 78 percent in the year to December 2025 and stood near 1,400 in July 2026 (devclass, 2026; PPC Land, 2026). Nobody decided this. Each developer rationally asked the model instead of the forum, and the forum was the model's source. Acemoglu, Kong and Ozdaglar (2026) model the same structure at the level of society, where agentic AI substitutes for the effort that produces general knowledge and can tip an economy into a state in which general knowledge vanishes despite high-quality personalised advice. We model the firm inside that market, and we are careful about which part of the mechanism belongs to the firm.

The model has two cohorts, seniors at full skill and a junior pool that enters at low skill and learns by practice, and a task distribution ordered by tacitness. AI supplies a skill-neutral floor on the codified part of the distribution and a smaller floor on the tacit part, where output quality also requires a skill threshold that only seniors meet. The tacit set has two parts. The larger is coverable by the vendor: the vendor's model codifies it over time whatever the firm does, because the vendor learns from everyone. The smaller is firm-specific, the clients, the legacy systems, the regulator's habits, the reasons a decision was taken in 2019, and it is codified only by the firm's own seniors working through the tool, and eroded by renewal whether or not anyone re-encodes it. Everything else is inherited from Caosun and Aral: the production function, the skill dynamics, the affine usage policy of a manager with a three-year horizon, and the five-region classification of deployments.

Four results follow. The first is a compression result. Adoption raises junior productivity on codified tasks by more than it raises senior productivity on tacit tasks, so required junior headcount falls first. Calibrated so that juniors gain 49 percent on codified tasks at full adoption and seniors 19 percent on tacit tasks, figures that bracket the field experiments, the model places the junior trough in 2027 in every calibration in which junior headcount reaches a trough, and reproduces the direction of the payroll gap of mid-2026. The second is a reversal result. Once adoption saturates, junior productivity stops rising and the frontier keeps moving, so tasks that migrate out of the tacit set are served by juniors with AI at a third of the loaded cost. The junior share of headcount rises from 58 percent to 66 percent by 2035 in the baseline and to 75 percent in a firm that captures its seniors' judgment; the senior stock halves in 2051 in the baseline, in 2037 in the capturing firm and between 2034 and 2039 under agentic or frontier calibrations, and whether required senior headcount halves at all is governed by a closed-form condition. The third is a formation result, and it is the one we did not expect. Senior formation requires practice on tasks the tool does not cover, that set is the complement of the frontier, and the frontier is rising; so the practice ground for the next senior shrinks in proportion to what the tool covers, and it is the same variable. The ladder inverts: a career that used to begin on the codified core and climb to the frontier must now begin at the frontier or not begin at all, and a firm that keeps juniors on uncovered work is the only firm still forming seniors. The fourth is a tax result. The output tax is conditional on the pipeline: it appears when the senior stock the pipeline still produces falls below what the uncovered set requires, which in the baseline happens in 2032, and it reaches 1.5 percent of task mass by 2035 and 9 percent by 2050. The firm's own residual is served first by the seniors it retains, so the firm-specific part stays below one percent until 2050 in the baseline. What the firm loses that nothing absorbs is its capacity on uncovered work, down 41 to 79 percent by 2035 to 2050, and its ability to rebuild its own frontier after a renewal shock.

The paper also prices the transfer. Using a receiving-end build-cost model, which values the capability a junior receives at what it cost to produce in one expert, we set the expert's loss, the junior's gain and the organisation's saving side by side for a reference team. Ninety-three percent of the build cost was financed by the worker, and the reproduction is priced to the firm as a seat. That is a statement about who paid; it is a distributional fact, and the paper does not make it an efficiency claim. The organisation's saving is large, immediate and visible; the measurable part of the tax is small and late; the larger part never appears in an output account. That asymmetry is the augmentation trap of Caosun and Aral at the level of the organisation, and it is why the reversal will be chosen rather than suffered.

Section 2 sets out the evidence on entry-level markets in 2026 and on the mechanism. Section 3 presents the model, the transfer assumption at two altitudes, and five propositions. Section 4 calibrates. Section 5 reports the replication, the fit to 2026, the reversal, the ledger and the tax. Section 6 walks through the objections in the order a reader will raise them, states what survives each, and gives testable predictions. Section 7 gives limitations and Section 8 concludes. Proofs and replication checks are in the Appendix.

## 2 Evidence

### 2.1 The entry-level market in 2026

Table 1 collects the indicators we use, with their sources and vintages. We report them as published and do not adjust them. Three features matter for the model.

| Indicator | Value | Period | Source |
|---|---|---|---|
| US recent-graduate unemployment rate | 5.6% | 2026 Q2 | Federal Reserve Bank of New York (2026) |
| US recent-graduate underemployment rate | 42% | 2026 Q2 | Federal Reserve Bank of New York (2026) |
| Employment, ages 22-25, most AI-exposed occupations, relative to less-exposed | −19% (−15% at Jul 2025 vintage) | Nov 2022 to Jun 2026 | Brynjolfsson et al. (2026), ADP payroll |
| Employment, ages 22-25, two most exposed quintiles (absolute) | −11%; +10% in three least-exposed quintiles | Nov 2022 to Jun 2026 | Brynjolfsson et al. (2026) |
| Software developers aged 22-25, employment vs late-2022 peak | about −20% | to Sep 2025 | Brynjolfsson et al. (2025) |
| Experienced workers in the same occupations | no comparable gap; flat or rising where AI augments | to Jun 2026 | Brynjolfsson et al. (2026) |
| US entry-level job postings, year on year | −7.5% | May 2026 | Indeed Hiring Lab (2026a) |
| US senior-level job postings, year on year | +14.7% | May 2026 | Indeed Hiring Lab (2026a) |
| Share of entry-level applications from workers with 10+ years' experience | 30% (largest group); 76% from 3+ years | May 2026 | Indeed Hiring Lab (2026b) |
| Class of 2026 hiring plans (median) | +5.6% (fall survey: −2.4%) | Spring 2026 | NACE (2026) |
| Entry-level postings requiring AI skills | 35% | 2026 | NACE (2026), via CNBC |
| Data and analytics postings mentioning AI | 45% | Dec 2025 | Indeed Hiring Lab (2026a) |
| Euro area youth (15-24) unemployment rate | 15.1% (+0.6 pp vs 2023 average) | 2026 Q1 | European Central Bank (2026) |

| Indicator | Value | Period | Source |
|---|---|---|---|
| Euro area youth employment, ICT / professional services / finance | −18.6% / −5.3% / −3.1% | 2023 Q1 to 2026 Q1 | European Central Bank (2026) |
| OECD youth (15-24) employment rate | fell; 55-64 supplied two-thirds of employment gains | 2024 Q1 to 2026 Q1 | OECD (2026) |
| Global youth unemployment rate; NEET rate | 12.4% (67 million); 20% (257 million) | 2025 | ILO (2026) |
| Northern America youth unemployment rate | 9.8% (8.3% in 2023) | 2025 | ILO (2026) |
| Youth employment in occupations highly susceptible to AI | 6.1% | 2025 | ILO (2026) |
| UK entry-level postings since Nov 2022 | about −30% | to 2025 | Adzuna, via techUK (2025) |
| UK graduate hiring; applications per employer | −8%; +44% | 2024/25 | Institute of Student Employers, via techUK (2025) |
| Netherlands unemployment, 15-24 (men / women) | 8.9% / 8.3%; all ages 4.0% | Jul 2026 | CBS (2026) |
| Netherlands, unemployment benefit claims under 27 | 46,889 (+8% in 2025; +16% in 2024) | end 2025 | UWV (2026) |
| US announced job cuts attributed to AI, year to date | 116,175 of 529,914 (22%); leading reason for five months | Jan to Aug 2026 | Challenger, Gray & Christmas (2026) |
| US college wage premium (log points) | 0.626 to 0.575 | 2022 to 2026 | Azar, Gine and Sanz-Espín (2026) |

Table 1: Entry-level and cohort indicators, 2025-2026. Figures are as published by the cited source; vintages differ and the series are not harmonised. The ECB and OECD attribute the bulk of youth weakness to cyclical cooling and long-run skill mismatch; the payroll evidence isolates an AI-exposure component.

First, the compression is concentrated and it is early-career. Employment of 22 to 25 year olds in the most exposed occupations fell 11 percent in absolute terms from November 2022 to June 2026 while the same age group in the least exposed occupations grew 10 percent, and the gap has widened at every data vintage (Brynjolfsson et al., 2026). The decline is confined to occupations where AI is used to automate; where it is used to augment, employment is flat or rising and the gains accrue to experienced workers. Adjustment runs through hiring, with little difference in compensation trends by age or exposure (Brynjolfsson et al., 2025). The macro series are consistent with this but cannot isolate it: youth unemployment has risen across eight of eleven global subregions since 2023 (ILO, 2026), and the ECB (2026) notes that existing evidence "does not yet allow this pattern to be attributed to generative AI or remote work specifically".

Second, the market has tilted toward seniority in the composition of demand and in the composition of supply. Senior postings rose while entry-level postings fell (Indeed Hiring Lab, 2026a), and experienced workers now dominate applications to entry-level roles (Indeed Hiring Lab, 2026b). The second fact is the one a cohort model must eventually explain, because a market that is tilting toward seniority in postings and toward experience in applications is a market in which the experienced are also being displaced, from somewhere.

Third, the roles that remain ask for more. Thirty-five percent of entry-level postings now require AI skills (NACE, 2026), 45 percent of data and analytics postings mentioned AI by December 2025 (Indeed Hiring Lab, 2026a), and the share of postings mentioning AI nearly doubled in a year (Handshake, 2026). Lightcast's skill disruption index found that the skills requested in the average US job changed 37 percent between 2016 and 2021 and that the pace was accelerating before language models arrived (Lightcast, 2022). The college wage premium has contracted from 0.626 to 0.575 log points since 2022, with task displacement in AI-exposed occupations accounting for about 28 percent of the fall and the series showing the first sustained negative relative demand growth for college labour since 1914 (Azar, Gine and Sanz-Espín, 2026). A cohort that is being asked to cover more of the task distribution at a lower premium is a cohort whose role is being redefined around the tool.

### 2.2 The mechanism

The experimental literature that Caosun and Aral organise with the (α, β) decomposition has a stable cross-sectional shape. Where a tool is deployed on codifiable work, the least experienced gain most: 34 percent for the least experienced customer-support agents against 14 percent on average (Brynjolfsson, Li and Raymond, 2025); 43 percent for below-median consultants against 17 percent for above-median (Dell'Acqua et al., 2026); a 55 percent rise in code output that was statistically significant only among junior developers, with seniors using the tool less rather than gaining less from it (Gambacorta et al., 2024). Where the work requires judgment, experienced workers use the tool differently and gain differently: high-tenure users of a frontier assistant are less likely to delegate through directive patterns and more likely to iterate on their own work (Anthropic, 2026), and experienced developers working with agents produce more aligned output and accept agent suggestions at higher rates while the gradient reverses for autocompletion (Sarkar, 2026). In the original model this is the difference between β below one on codified tasks and β above one on tacit tasks.

What the cross-section cannot show is the direction of the frontier. The payroll evidence identifies the boundary between the two regimes as the boundary between codified and tacit knowledge: employment has declined among young workers in occupations that rely on "formal, standardized, documented knowledge that can be taught through education, textbooks, or written procedures", and increased among experienced workers in occupations relying on knowledge "acquired through practice, mentorship, and repeated exposure to real situations" (Brynjolfsson et al., 2026). The boundary is not fixed. It moves as tacit knowledge becomes documented, and the documentation now happens inside the tool. Reinforcement from expert feedback, enterprise fine-tuning on corrected drafts, agent traces reviewed and approved by senior staff, and workflow capture are all mechanisms by which the exercise of judgment through an AI system becomes training signal for that system. We do not have a measured rate for this process, and we treat its rate as the model's least identified parameter. We do have a measured consequence at the level of the public commons. Stack Overflow, the forum that supplied much of the coding knowledge on which the first models were trained, received more than 200,000 new questions a month at its 2014 peak. The decline began before language models, as search engines and documentation improved, and it accelerated after November 2022: December 2025 recorded 3,862 questions, 78 percent fewer than a year earlier, and July 2026 about 1,400 (devclass, 2026; PPC Land, 2026). The series is the source population of a shared model going quiet within three years of the model's release, without anyone deciding that it should, and it is the observation on which the market-level form of our assumption rests. The firm-level analogue is narrower: the seniors who supply the training signal for the firm's own residual are the seniors whose tasks are being codified.

The displacement half of Assumption 1 has been measured directly in two settings, one on each cohort.

On the senior side, Budzyń et al. (2025) compared unassisted colonoscopy at four Polish endoscopy centres in the three months before a computer-aided detection system entered routine use with the three months after, holding the procedures unassisted in both windows. Across 1,443 non-AI colonoscopies, the adenoma detection rate fell from 28.4 to 22.4 percent, an absolute difference of six percentage points (95% CI −10.5 to −1.6, $p = 0.009$), with exposure to AI an independent predictor in multivariable analysis (OR 0.69, 95% CI 0.53–0.89). The nineteen endoscopists had each performed more than 2,000 colonoscopies. This is skill at or near $\bar{S}$ eroding under interior usage over three months, which is the law of equation (2) applied to seniors and observed in the field. The study is retrospective and observational, drawn as a nested cohort from the ACCEPT trial whose selection into the non-AI arm the manuscript does not fully describe, and the authors flag selection bias and residual confounding; we treat it as one measured case rather than an identified effect. The three-month window is also shorter than the feedback channels available to an endoscopist, interval cancers and follow-up detection among them, so the design cannot distinguish erosion from a behavioural carryover that would reverse once a miss surfaced and was traced. It does not observe a correction phase and it cannot rule one out.[1]

On the junior side, Bastani et al. (2025) randomised nearly a thousand high-school students across two deployments of the same model and a control. Unrestricted access raised performance on assisted practice problems by 48 percent and lowered performance on a subsequent unassisted examination by 17 percent against control. A second deployment, restricted to withhold solutions and prompt the student's own working, raised assisted practice performance by 127 percent and left unassisted performance intact. The population is students and the horizon is one course, so the experiment does not identify κ for a workforce. What it identifies is that the two arms differ in deployment design and not in model capability, which is the empirical content of the claim in

[1] We owe this reading of the study's window to Jeremy McEntire, in correspondence, September 2026.

Section 3.3 that λ, like β, is a property of the workflow, and of Proposition 4's dependence on where the junior deployment sits relative to the automation boundary.

Why does AI-assisted work fail to build expertise? Experimental evidence locates one mechanism in individual judgement. Across five experiments (N = 3,132; four pre-registered, one direct replication), Marcoccia, Quattrociocchi and Capraro (2026) show that access to AI advice nearly eliminates people's willingness to say "I don't know", even though the questions were engineered so that the advice was almost always wrong. In Study 4, where advice appeared automatically rather than on request, judgement suspension fell from 35 percent without AI to 1 percent with it. Pooled across the unincentivised conditions, participants answered more questions but were correct about a third as often (9.2 against 27.5 percent), while mean confidence rose from 29.6 to 75.9 on a hundred-point scale. Incentivising accuracy raised both suspension and accuracy, but did so independently of AI access rather than by blunting its effect, and suspension remained far below the level observed when AI was unavailable.

In our two-cohort model, expertise forms through unassisted practice on work the tool cannot do. If a fluent answer pre-empts the worker's own judgement, assisted practice does not count towards that formation. The authors conclude that "people can recruit their own judgement when motivated, so what AI changes is willingness, not capacity". This is consistent with a model that concerns people who begin their careers with the tool already in place. For them, a persistent reduction in the willingness to exercise judgement becomes, over time, capacity that never forms.

The authors flag the boundary condition of AI reliability as untested: their questions were engineered so the advice was almost always wrong, and whether suspension collapses to the same degree when advice is usually correct is unknown. That case is the one our model addresses. When advice is usually right, deferring to it is rational in every single decision, yet the same deferral removes the practice through which expertise forms. Deference that is rational case by case can therefore be costly over a career.

Two further facts constrain the model. Firms have begun to redesign entry-level roles around the tool rather than eliminate them. IBM announced it would triple entry-level hiring in the United States in 2026, for "all these jobs that we're being told AI can do", with job descriptions shifted away from coding toward customer-facing work and explicitly to secure a pipeline into senior roles (TechCrunch, 2026). AI was the leading stated reason for announced US job cuts for five consecutive months to July 2026, 116,175 cuts year to date, though total announced cuts were 41 percent below 2025 and announced hiring 37 percent above (Challenger, Gray & Christmas, 2026). The picture is one of reallocation across cohorts rather than net destruction, which is what a two-cohort model with a moving frontier produces.

## 3 Model

### 3.1 Inherited structure

We keep the production function and skill dynamics of Caosun and Aral (2026) unchanged. A worker with skill $S$ using AI at intensity $u \in [0, 1]$ produces

$$p(S, u) = (1 - u) S + [\alpha + \beta S - \gamma u] u, \quad (1)$$

where $\alpha$ is the skill-neutral productivity of the tool, $\beta S$ the knowledge-complementary productivity that scales with the worker's judgment, and $\gamma > 0$ enforces diminishing returns to delegation. Skill follows the learning-forgetting law

$$dS/dt = \kappa [ \bar{S} (1 - u) - S ], \quad (2)$$

so that practice without AI accumulates skill toward the potential $\bar{S}$ and offloading displaces it. A decision-maker with discount rate $\delta$ chooses $u$ to maximise discounted output. Lemma 1 of the original paper gives a quadratic value function $V(S) = aS^2 + bS + c$ and an affine usage policy $u^\star(S) = u_0 + u_1 S$, with the coefficients in closed form (their equations EC.6 to EC.9; reproduced in Appendix A). We use the manager's policy throughout, with $\delta_F = 0.33$ for a three-year evaluation horizon, and report the worker's horizon $\delta_W = 0.10$ as a sensitivity. The five-region classification of deployments in $(\alpha, \beta)$ space, with its adoption boundary $C_0$, automation boundary $C_1$ and break-even locus $B$, is inherited and replicated in Section 5.1.

### 3.2 Tasks, cohorts and the codified frontier

The firm must serve a unit mass of tasks indexed by tacitness $x \in [0, 1]$. A task is *codified* if $x \leq c_t$ and *tacit* otherwise; $c_t$ is the *codified frontier*, the share of the firm's task mass on which the tool supplies its full skill-neutral floor. On codified tasks the deployment has parameters ($\alpha A_t$, $\beta_c$) with $\beta_c < 1$: the tool handles most of the work and judgment adds little, the skill-levelling regime of Section 2.5 of the original paper. On tacit tasks the deployment has parameters ($\theta\alpha A_t$, $\beta_\tau$) with $0 < \theta < 1$ and $\beta_\tau > 1$: the tool supplies a smaller floor and judgment shapes the quality of its output, the skill-complementary regime of Section 2.4. The term $A_t \in [0, 1]$ is adoption diffusion, a logistic path from zero in November 2022 that scales the effective α.

The tacit set has two parts. A share φ of the task mass is *firm-specific*: tasks whose tacit content is particular to this firm and cannot be learned from any other firm's output. The remaining share 1 − φ is *vendor-coverable*: tasks whose tacit content is common across firms and will be codified by the vendor's model over time whatever this firm does. We track a frontier for each, $c_{f,t}$ and $c_{v,t}$, and the aggregate frontier is $c_t = \varphi c_{f,t} + (1 - \varphi)c_{v,t}$.

Tacit tasks carry a quality requirement: output at full quality requires skill at or above a threshold $S_{min}$. A worker with $S < S_{min}$ assigned to a tacit task delivers a fraction $S/S_{min}$ of its value. This is the model's definition of seniority. A *senior* is a worker with $S \geq S_{min}$; we set senior skill at the potential, $S_H = \bar{S} = 1$, and hold it there. This is a simplification and not a claim: Section 2.2 reports field evidence that senior skill erodes under interior usage on the same law as the junior pool's, so the assumption suppresses a channel the paper elsewhere treats as real. It is made to keep the cohort structure tractable, and it is not neutral. Eroded seniors serve less tacit mass each, so required senior headcount would fall more slowly and the halving dates in Table 3 would be later; they also fall below $S_{min}$, so a quality shortfall would appear on tacit work that the model as specified does not price. The tax results are therefore understated and the reversal dates overstated, and a reader should carry both directions. A *junior* is a member of a pool that enters at skill $S_e$ and turns over at rate ρ through promotion and attrition, so that the pool's mean skill obeys

$dS_L/dt = \kappa [ \bar{S} (1 - u_L) - S_L ] + \rho ( S_e - S_L )$. (3)

Juniors serve codified tasks and seniors serve tacit tasks; this is the task assignment that minimises cost under the quality requirement, and it is the pre-AI pyramid. Each cohort's usage is set by the manager's affine policy evaluated at its own task parameters and skill: $u_H = u^\star(S_H; \theta\alpha A_t, \beta_\tau)$ and $u_L = u^\star(S_L; \alpha A_t, \beta_c)$. Per-worker output on each task class follows from (1), and the headcount required to serve unit demand is

$n_{L,t} = c_t / p_{L,t}$, $n_{H,t}^{req} = (1 - c_t) / p_{H,t}$. (4)

Cohorts are paid $w_L$ and $w_H$ with a common employer loading, plus a seat cost per user. Wages are held fixed. The payroll evidence shows adjustment through employment and not through compensation (Brynjolfsson et al., 2025), and a model with fixed wages and free quantities is the one that evidence selects.

### 3.3 The transfer, at two altitudes

The extension rests on one assumption, which we state first at the altitude at which the evidence supports it and then at the altitude at which the firm's decisions matter.

**Assumption 1 (Capture).** *The tacit capability a tool supplies on a class of tasks is produced by unassisted practice with stakes on that class, and use of the tool on that class displaces the practice that produced it. At the level of the market, the shared model's frontier on vendor-coverable work advances at a rate set by the practice of the population that still performs that work unaided, and it is exogenous to any one firm. At the level of the firm, the frontier on firm-specific work advances only through the firm's own seniors' usage:*

$dc_f/dt = \lambda\, n_{H,t}\, u_{H,t}\, S_H (1 - c_{f,t}) - \mu\, c_{f,t}$, $dc_v/dt = \lambda_v (1 - c_{v,t}) - \mu\, c_{v,t}$, (5)

*where $\lambda \geq 0$ is the firm's capture rate, $\lambda_v \geq 0$ the vendor's codification rate, and $\mu \geq 0$ the task-renewal rate.*

The firm-level source term is the offloading term of equation (2) summed over seniors and applied to the firm-specific tacit mass. In the original model, $\kappa u_H S_H$ is the rate at which a senior's practice is displaced by delegation; here, $\lambda u_H S_H$ is the rate at which the judgment exercised through the tool becomes part of what the tool can do without the senior. The two are the same act observed from two sides. The factor $(1 - c_f)$ restricts the source to tasks not yet codified, and μ removes codified tasks as the task distribution renews through new products, regulation and technology; renewal erodes the firm-specific codified stock whether or not anyone re-encodes it, because documentation of a firm's own practice decays when the practice changes. The vendor's frontier has no senior in its source term. This is the free-rider fact: one firm's seniors are a rounding error against the vendor's corpus, and the firm that releases them changes nothing about the tool it rents. The vendor's rate $\lambda_v$ is set by the market-level version of the assumption, by whoever still performs the common tasks unaided, and Section 2.2 gives the one measured case of that population going quiet.

The capture rate λ is a property of the deployment rather than of the technology: a firm that logs and learns from expert corrections, that fine-tunes on reviewed output, or that lets agents act under senior approval has a higher λ than one that does not, in the same way that β in the original model is a property of workflow design. Each frontier has a steady state. The vendor's is $c_{v,\infty} = \lambda_v/(\lambda_v + \mu)$. The firm's satisfies $\lambda n_H u_H S_H(1 - c_f) = \mu c_f$ with the senior stock at the total requirement $(1 - c)/p_H$ and the vendor's frontier at its own steady state, in the approximation that the stock tracks requirement, which holds while the stock exceeds it. Writing $x = 1 - c_{f,\infty}$ and $k \equiv \lambda u_H S_H/p_H$, this is the quadratic $k\varphi x^2 + [k(1 - \varphi)(1 - c_{v,\infty}) + \mu]x - \mu = 0$, whose admissible root is

$$1 - c_{f,\infty} = [ -b + (b^2 + 4k\varphi\mu)^{1/2} ] / (2k\varphi), \quad b \equiv k (1 - \varphi)(1 - c_{v,\infty}) + \mu, \quad k \equiv \lambda u_H S_H / p_H. \quad (6)$$

At λ = 1 and the baseline values of the other parameters the root gives $c_{f,\infty} = 0.675$, against a simulated peak of 0.690; at λ = 0 it is zero, because without capture the firm-specific frontier decays at rate μ toward zero.

Because the firm-level source term is proportional to the senior headcount, the firm-specific frontier cannot advance without seniors and cannot be maintained against renewal without them; when seniors are gone, $c_f$ decays at rate μ toward zero while $c_v$ continues to rise. This asymmetry, a vendor frontier that moves regardless and a firm frontier that moves only with the firm's own practice, is what the firm-level results turn on.

### 3.4 Senior stock and the pipeline

The stock of seniors on payroll, $n_{H,t}$, is replenished by juniors who cross $S_{min}$ and depleted by retirement at rate $r$ and by release when the stock exceeds requirement. An earlier draft measured replenishment by the practice flow of equation (2) on the junior's own tasks. That is the wrong quantity. $S_{min}$ is competence on tacit tasks, tasks on which the tool's output cannot be trusted, and competence there is built by doing them; practice on codified tasks builds fluency the tool already supplies. Senior formation therefore requires practice on *uncovered* work, and the practice ground is the complement of the frontier. We define the *pipeline index* $\Pi_t$ as unassisted junior practice weighted by the uncovered task mass, relative to its pre-AI value:

$$\Pi_t = n_{L,t} (1 - u_{L,t}) (1 - c_t) / [ n_{L,0} (1 - c_0) ], \quad (7)$$

and let seniors be produced at rate $\pi_0\Pi_t$, with $\pi_0 = r\, n_{H,0}$ so that the pre-AI pipeline exactly offsets retirement. Surplus seniors are released at rate ξ per year, which allows for hoarding and attrition-first adjustment; the senior stock then follows

$$dn_H/dt = \pi_0 \Pi_t - r\, n_{H,t} - \xi \max\{0, n_{H,t} - n_{H,t}^{req}\}. \quad (8)$$

The pipeline index falls for two reasons that the earlier draft conflated: juniors delegate, which is the original model's erosion, and the uncovered set shrinks, which is the frontier. It is zero when juniors delegate fully, and it is decreasing in $c$ at every level of usage. The second channel is the one that produces the ladder result in Section 3.6. We also allow the firm a lever the pre-AI pyramid did not need: a share $s$ of uncovered work assigned to juniors unaided, at quality $S_L/S_{min}$. This is *frontier apprenticeship*, and it is the only way, in the model, to form a senior once the codified core belongs to the tool.

### 3.5 The adaptability tax

When the senior stock falls below requirement, tacit tasks are served by juniors at fractional quality or not at all. Seniors serve the firm-specific residual first, because they are the only workers who can, and vendor-coverable tacit work second.

**Definition 1 (Adaptability tax).** *The adaptability tax at t is the share of the firm's task mass lost on tacit tasks that no senior is available to serve, net of the quality juniors can supply, reported separately for firm-specific and vendor-coverable work:*

$$\tau_{f,t} = \max\{0, \varphi(1 - c_{f,t}) - n_{H,t}p_{H,t}\} \cdot q_t, \quad \tau_{v,t} = \max\{0, (1 - \varphi)(1 - c_{v,t}) - \max\{0, n_{H,t}p_{H,t} - \varphi(1 - c_{f,t})\}\} \cdot q_t, \quad (9)$$

with $q_t = 1 - \min\{S_{L,t}/S_{min}, 1\}$. The distinction matters because $\tau_v$ is transient: the vendor's frontier keeps rising and absorbs the work eventually. $\tau_f$ is the firm's own. Two further measures capture what τ does not. The *frontier capacity* $\mathcal{I}_t = n_{H,t}(1 - c_t)$ is the senior effort available on uncovered work; it is the quantity that advances the firm-specific frontier in (5) and the firm's capacity to solve problems the tool has not seen, which is the operational content of innovation in this model. The *rebuild time* $T_{R,t} = (1/\kappa) \ln[(\bar{S} - S_{L,t})/(\bar{S} - S_{min})]$ is the number of years of unassisted practice needed to produce one senior from the junior pool at its current mean skill; it follows from (2) at $u = 0$.

### 3.6 Results

Let $g_L$ and $g_H$ denote the proportional productivity gains at full adoption on codified and tacit tasks, $g_L = p_L/S_{L,0} - 1$ and $g_H = p_H/S_H - 1$, evaluated at the manager's policy.

**Proposition 1 (Junior compression).** *Hold c fixed. During adoption diffusion, required junior headcount falls monotonically toward $c/p_L$ and required senior headcount toward $(1 - c)/p_H$. At full adoption the proportional declines are $g_L/(1 + g_L)$ and $g_H/(1 + g_H)$. Junior demand falls by more than senior demand if and only if $g_L > g_H$, which holds whenever $\theta < 1$ and $\beta_c < 1 < \beta_\tau$ at interior usage.*

The proof is immediate from (4). The condition $g_L > g_H$ is the cross-sectional fact of Section 2.2, that the least experienced gain most from tools deployed on codifiable work, restated as a condition on task parameters. In a firm where juniors do the codified work, a tool that raises junior productivity by more than senior productivity reduces junior demand by more than senior demand, whatever it does to the frontier.

**Proposition 2 (Cohort reversal).** *Suppose $c_0 < c_\infty = \varphi c_{f,\infty} + (1 - \varphi)c_{v,\infty}$. Then required junior headcount reaches a minimum after which it increases, required senior headcount decreases throughout, and the junior share of required headcount rises monotonically toward $c_\infty p_H/[c_\infty p_H + (1 - c_\infty)p_L]$. The turning point occurs when the proportional rate of frontier advance exceeds the proportional rate of junior productivity growth, $\dot{c}/c > \dot{p}_L/p_L$, which is guaranteed once adoption saturates. The reversal occurs even at $\lambda = 0$, driven by the vendor's frontier alone.*

The last sentence is the free-rider correction. The reversal is not something the firm does to itself through capture; it is what the vendor's improving model does to every firm's pyramid, and the firm's own λ changes its timing at the margin. Whether junior headcount recovers its 2022 level in absolute terms depends on whether frontier advance outruns junior productivity; Section 5 reports that it does in the capturing firm and does not in the baseline or under the frontier calibration, where each junior with AI covers so much more of the codified set that fewer are needed even as the set grows.

**Proposition 3 (Senior halving).** *Required senior headcount falls to half its pre-AI level if and only if $c_\infty \geq c^\dagger \equiv 1 - (1 - c_0)p_H/(2p_{H,0})$. The halving date is decreasing in $\lambda_v$, in $\lambda$ and in $u_H$, and increasing in $\mu$. Under the manager's policy $u_H$ is increasing in $\delta_F$, so a shorter evaluation horizon brings the date forward through the firm-specific channel.*

The condition compares the aggregate frontier's steady state, which the vendor mostly sets, with the level at which the tacit set is small enough to be served by half the seniors. The comparative static in $\delta_F$ is Proposition 5 of the original paper applied to the source term of (5).

**Proposition 4 (The tax is conditional on the pipeline).** *If the junior deployment $(\alpha A_t, \beta_c)$ lies in the interior of the adoption region for the junior pool, so that $u_L < 1$, the pipeline index is bounded away from zero as long as $c < 1$ and the senior stock converges to $\Pi_\infty n_{H,0}$. The stock covers requirement, and $\tau_f = \tau_v = 0$ in steady state, if and only if the pipeline condition $\Pi_\infty n_{H,0} \geq (1 - c_\infty)/p_H$ holds; when it fails, $\tau_v$ becomes positive at the date at which the stock falls below requirement, and $\tau_f$ becomes positive only if the stock falls below $\varphi(1 - c_f)/p_H$. If the junior deployment lies in the automation region, so that $u_L = 1$, the pipeline index is zero, the condition fails, the senior stock decays at rate r, $\tau_v$ becomes positive from the date at which $r\, n_H$ exceeds the release of surplus seniors and is absorbed as $c_v \to c_{v,\infty}$, and $\tau_f$ becomes positive when the stock falls below $\varphi(1 - c_f)/p_H$, where $c_f \to 0$ at rate $\mu$ once the source term vanishes. Seniors serving the residual first, $\tau_f$ is the last component of the tax to appear and the only one that does not resolve.*

Proposition 4 locates the tax on the original paper's region map and, after the free-rider correction, locates it inside the firm. A junior deployment that stays in Region II or IV leaves the practice channel partly open; a deployment that reaches Region III or V closes it. An interior deployment leaves the channel open but does not by itself keep the stock at requirement: in the baseline $\Pi_\infty = 0.37$ against a requirement of 0.74 of the 2022 stock, so the condition fails and the stock falls below requirement in 2032. Only frontier apprenticeship (Section 5.4) brings the stock back to within two percent of requirement. And the part of the tax that the firm cannot rent its way out of is the smallest part, the firm-specific residual, which is why it is the last to appear and the easiest to ignore.

**Proposition 5 (Ladder inversion).** *Senior formation requires unassisted practice on uncovered work, whose mass is $1 - c_t$. Under Assumption 1, $c_t$ is non-decreasing after adoption, so the practice ground for senior formation is non-increasing and converges to $1 - c_\infty$. Holding the junior pool's task assignment at the pre-AI pyramid, the pipeline index at $u_L = 0$ equals $c_t(1 - c_t)/[c_0(1 - c_0)]$: the shrinking ground is partly offset by the* *larger junior headcount the codified mass requires, and the index falls below one once $c_t$ passes one half, toward $c_\infty(1 - c_\infty)/[c_0(1 - c_0)]$. The pipeline can be held above that floor only by assigning juniors to uncovered work unaided ($s > 0$), which raises the required junior headcount, lowers the required senior headcount, and prices the quality shortfall on that work into current output at $(1 - S_L/S_{min})$ per unit.*

Proposition 5 is the paper's sharpest and least comfortable result. The tasks that remain uncovered are, by construction, the hardest tasks in the firm: firm-specific judgment and genuinely novel work. A career used to begin on the codified core and climb toward those tasks. The codified core is now the tool's. So the only entry-level work that still forms an expert is work above the coverage line, and the firm that assigns it to juniors, and places answerability for it with them, paying the quality cost in current output, is the only firm still forming seniors of its own. The proposition concerns the firm's allocation of work and of consequence. It is not a claim that an individual can no longer become expert: formation predates the institutions that now house it, and a practitioner who carries the consequence of a decision still learns from it. What the firm controls, and what the model tracks, is whether that consequence reaches the junior at all. The apprenticeship model of every profession that depends on sustained unassisted practice during formative years, the residency, the associate track, the engineering apprenticeship, is inverted: the formative years must now be spent on the hardest work rather than the easiest, and the firm must pay for that in visible quality rather than receive it in cheap labour.[2]

## 4 Calibration

Table 2 lists every parameter, its value and its provenance. Parameters inherited from Caosun and Aral keep their illustrative values where the original paper uses them; parameters we add are set to published moments where a moment exists and flagged as assumptions where it does not.

| Parameter | Value | Provenance |
|---|---|---|
| $\gamma$ (diminishing returns) | 1.0 | Caosun & Aral, Figures 1-3 |
| $\kappa$ (learning/ forgetting rate) | 0.2 | Between the original paper's 0.3 (Figs 1-4) and 0.1 ("a decade to build", Fig 5); gives 4.9 years from the pre-AI junior pool to senior |
| $\delta_F$, $\delta_W$ (discount rates) | 0.33, 0.10 | Caosun & Aral, Section 2.7: three-year manager, ten-year worker |
| $\bar{S}$, $S_H$ | 1.0 | Normalisation, as in the original |
| $\beta_c$ (codified tasks) | 0.5 | Caosun & Aral, Figure 3 skill-levelling case |
| $\beta_\tau$ (tacit tasks) | 1.5 | Caosun & Aral, Figure 3 skill-complementary case |
| $\alpha$ (codified floor, full adoption) | 1.6 (1.4; 2.0) | Calibrated: junior gain on codified tasks +49% at the pre-AI pool skill. Brackets Brynjolfsson et al. (2025) +34% least experienced; Dell'Acqua et al. (2026) +43% below median; Gambacorta et al. (2024) +55% juniors. $\alpha$=1.4 reproduces +33%; $\alpha$=2.0 reproduces the June 2026 relative employment gap of −19% |
| $\theta$ (tacit floor as share of $\alpha$) | 0.3 | Calibrated: senior gain on tacit tasks +19%. Dell'Acqua et al. (2026) +17% above median; Brynjolfsson et al. (2025) +14% average |
| $S_e$ (entry skill) | 0.2 | Assumption |
| $S_{min}$ (senior threshold) | 0.9 | Assumption; with $\kappa$=0.2 the pre-AI pipeline takes 4.9 years from pool mean and 10.4 years from entry |

[2] An earlier version of this discussion was phrased as a claim about careers rather than about firms. Jeremy McEntire objected, in correspondence, that senior formation predates teachers and reverts to learning from consequence when instruction is withdrawn, so the defensible claim concerns the assignment of work and of answerability. We accept the objection and have restated the proposition accordingly.

| Parameter | Value | Provenance |
|---|---|---|
| ρ (junior pool turnover) | 0.10 | Assumption: ten-year mean tenure in the junior pool |
| r (senior retirement rate) | 0.05 | Assumption: twenty-year senior careers |
| $c_0$ (codified share, 2022) | 0.50 | Assumption; implies a pre-AI pyramid of 0.68 juniors per 0.50 seniors, junior share of headcount 58% |
| φ (firm-specific share of the tacit set) | 0.4 (0.2; 0.6) | Assumption; the share of tacit work no vendor can learn from other firms' output |
| $\lambda_v$ (vendor codification rate) | 0.10 (0.20) | Assumption; exogenous to the firm, set by the market-level form of Assumption 1. Implies c_v,∞ = 0.77 |
| λ (firm capture rate) | 0 (0.25; 0.5; 1.0; 2.0) | Not identified. The baseline is the free-riding firm, which Section 5.7 argues is the 2026 prior for most firms; λ = 1 is reported throughout as the capturing firm. Section 5.3 reports the halving dates over the (λ, μ) plane |
| s (frontier apprenticeship) | 0 (0.2) | Share of uncovered work assigned to juniors unaided; a policy lever, Proposition 5 |
| μ (task-renewal rate) | 0.03 (0.015; 0.06) | Assumption; Lightcast (2022) reports 37% of requested skills changed over five years, an upper bound on renewal of the codified stock |
| $A_t$ (adoption diffusion) | logistic, midpoint mid-2025, A(mid-2026)=0.86 | Assumption consistent with survey evidence of majority workplace use by 2025-26 |
| ξ (release rate of surplus seniors) | 0.2 per year | Assumption: attrition-first adjustment; consistent with flat senior employment through mid-2026 |
| $w_H$, $w_L$, loading, seat | €85,000; €36,000; 1.3; €600 | Companion workbook (Dutch senior professional band, graduate entry, employer loading, licence); wage ratio 2.36 |

Table 2: Parameters. Values in parentheses are the alternative calibrations reported in Section 5. The reference year is November 2022 (t = 0).

Three calibration choices deserve comment. The first is α. The productivity experiments were run with 2023 to 2024 tools; the payroll gap of June 2026 reflects 2025 to 2026 tools. Calibrating to the experiments gives α = 1.4 and a junior gain of 33 percent; calibrating to the employment gap gives α = 2.0 and a junior gain of 89 percent on codified tasks. We use α = 1.6 as the baseline and report both ends, noting that the original paper's Section 3.5 anticipates that α rises as models improve. The second is the split between $\lambda_v$ and λ. No published series measures either rate, and we do not construct one; we set the vendor's rate so that the vendor frontier alone moves the aggregate frontier at roughly the pace an earlier draft attributed to the firm, we set the baseline capture rate at zero because that is where Section 5.7 places most firms in 2026, and we report the halving dates over the (λ, μ) plane and a capturing firm at λ = 1 so that a reader who holds a different view can read off the date it implies. The third is φ. The firm-specific share of tacit work is not measured anywhere we know of; we set it at 0.4 and report the results over 0.2 to 0.6. Without capture it is the share of the task mass that can never be codified, and Section 5.7 shows that the pipeline is insensitive to it and the tax is not.

## 5 Results

### 5.1 Replication

Before extending the model we reproduce it. Our implementation of the closed forms EC.6 to EC.9 returns the illustrative numbers of Section 2.7 of the original paper: under $(\alpha, \beta, \kappa, \bar{S}) = (0.8, 1.3, 0.3, 1)$, a manager with $\delta = 0.33$ sets steady-state usage $\hat{u} = 0.255$ and a worker with $\delta = 0.10$ sets $\hat{u} = 0.139$, with steady-state skill 0.745 against 0.861, matching the reported 0.26, 0.14, 0.75 and 0.86. At $\beta = 1$ the adoption and break-even thresholds are $\alpha_0 = 0.75$ and $\alpha_1 = 1.25$ as in their Figure 1, and the steady states of their Figure 3 are recovered to three decimal places. Figure 1 reproduces the five-region map of their Figure 4, including the small Region III wedge near $(\alpha, \beta) = (1.15, 1.0)$.

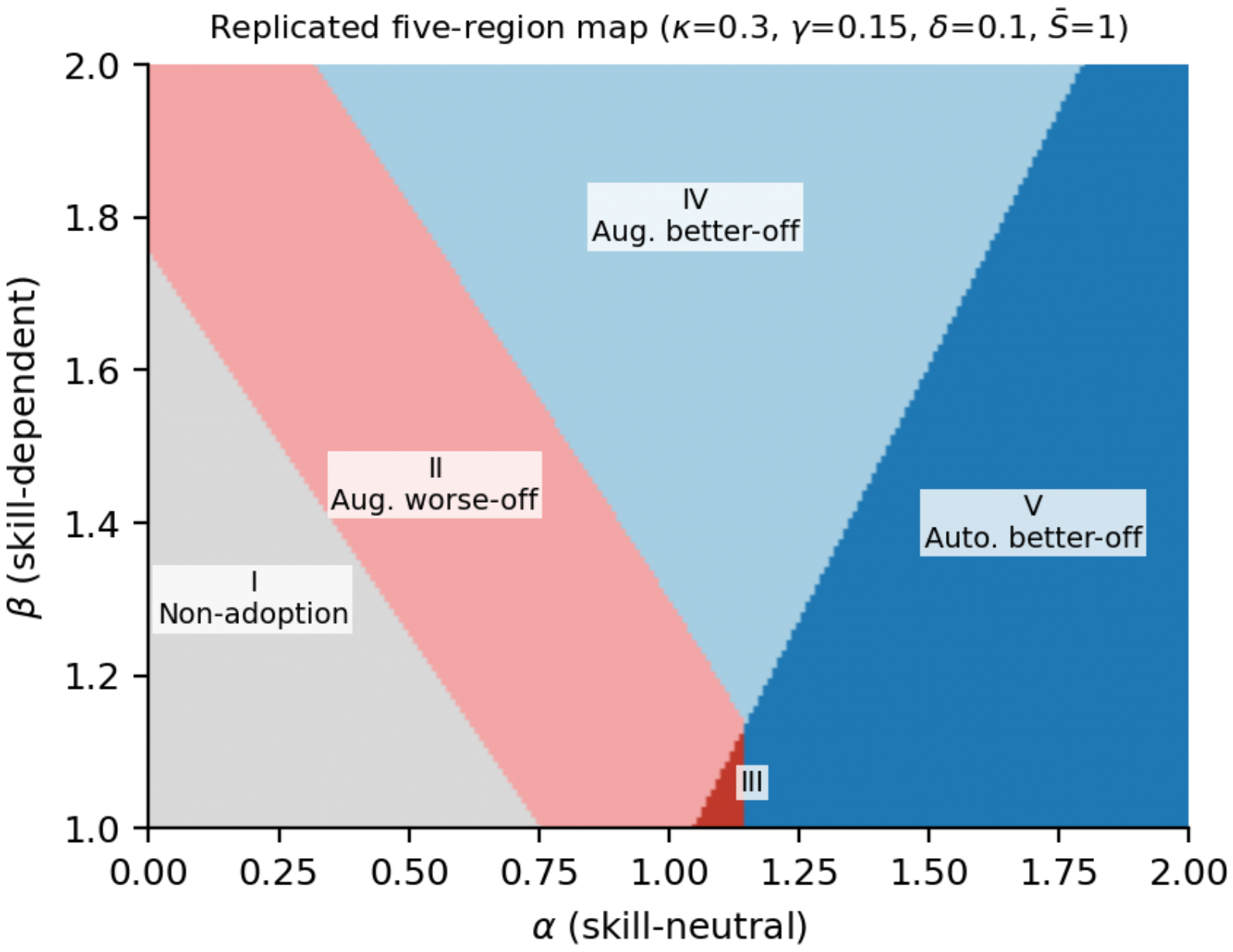

Figure 1: Replicated five-region map of Caosun and Aral (2026), Figure 4, at $\gamma = 0.15$, $\kappa = 0.3$, $\delta = 0.1$, $\bar{S} = 1$. Regions are computed from the closed-form value function and steady state (Appendix A).

**5.2 Phase one: compression, 2023 to 2027**

Figure 2 shows the baseline path. By early 2026 junior usage on codified tasks has risen to 0.25 and senior usage on tacit tasks to 0.22; by mid-2026 the aggregate frontier has moved from 0.50 to 0.54, all of it the vendor's, and required junior headcount is 15 percent below its November 2022 level. The payroll data show an 11 percent absolute decline for the exposed young cohort (Brynjolfsson et al., 2026). The frontier calibration ($\alpha = 2.0$) gives a junior decline of 30 percent and a gap against the senior stock of 20 percent, against a published relative gap of 19 percent; the experiment calibration ($\alpha = 1.4$) gives a junior decline of 8 percent and a relative gap of 1 percent. The model does not fit these moments by construction: $\alpha$ is set to the productivity experiments, the adoption curve is imposed, and the employment outcome follows from Proposition 1. One discrepancy is worth stating. The model's required senior headcount is 17 to 31 percent below its 2022 level by mid-2026 across calibrations, because senior productivity rises and the vendor frontier moves, while the senior stock, released at the attrition-first rate, is 7 to 14 percent down; the payroll data show senior employment flat or rising in the same occupations. The gap is the difference between required headcount at fixed demand and actual headcount under labour hoarding and demand growth, and it is why the relative gap in the baseline, 8 percent, is smaller than in the data.

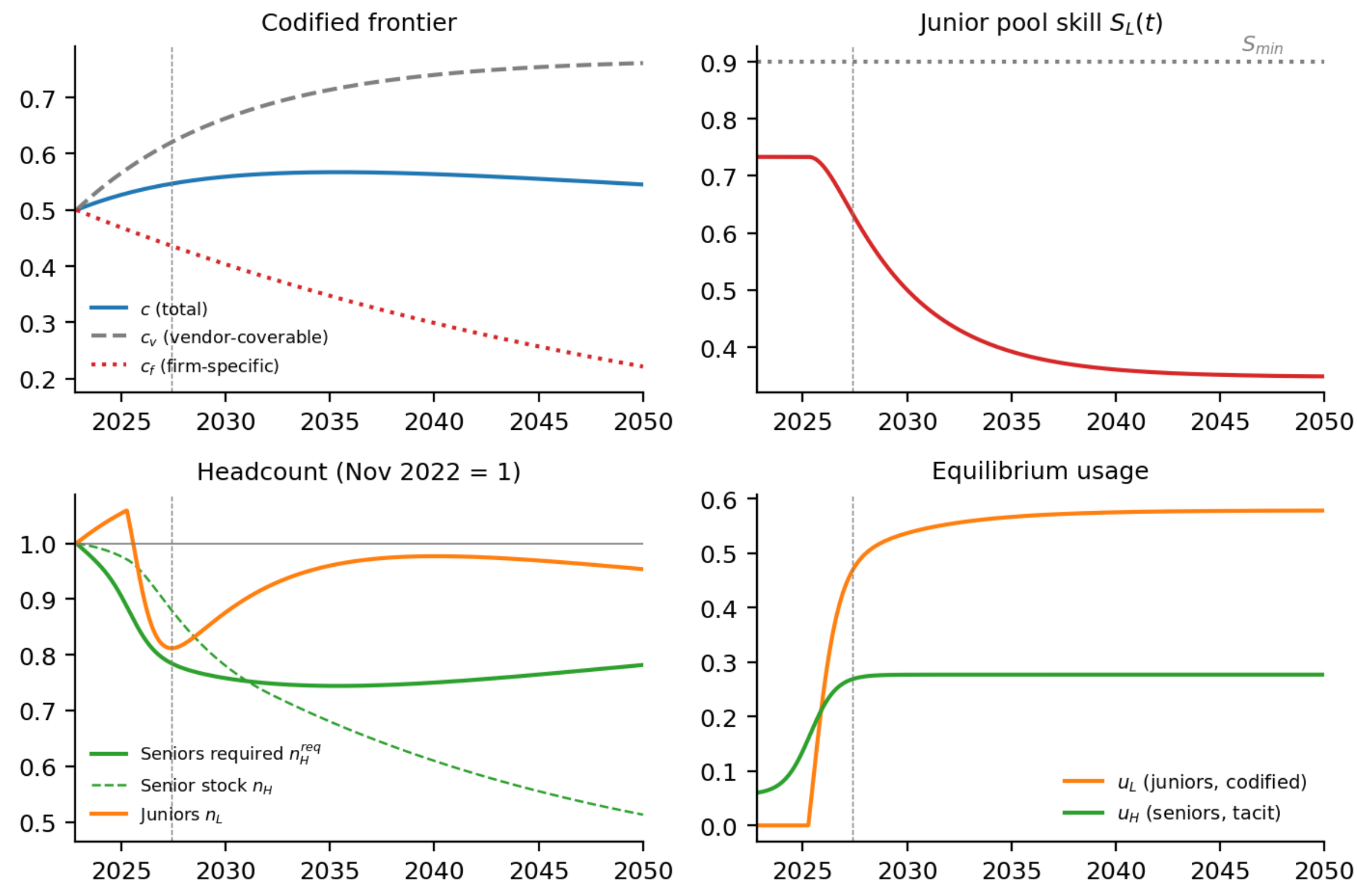

Figure 2: Baseline dynamics, November 2022 to 2050 ($\alpha = 1.6$, $\theta = 0.3$, $\varphi = 0.4$, $\lambda_v = 0.10$, $\lambda = 0$, $\mu = 0.03$). Top left: the aggregate frontier and its two components; the vendor frontier converges to 0.77 and the firm-specific frontier, with no capture, decays at the renewal rate. Top right: the junior pool's mean skill falls from 0.73 to 0.35 by 2050 and never approaches $S_{min} = 0.9$. Bottom left: junior headcount, required senior headcount and the senior stock; the junior trough is in the second quarter of 2027 and the stock crosses below requirement in 2032. Bottom right: the manager's equilibrium usage by cohort. The dashed line marks the junior trough.

The junior trough falls in 2027 in every calibration in Table 3 in which junior headcount reaches a trough, in the second or third quarter of that year in most of them, because its timing is set by the imposed adoption curve and not by the frontier; the trough date is therefore an input of the calibration, and what the model adds is its depth and what follows it. Two scenarios, fast renewal and a high firm-specific share, have no trough because junior headcount falls throughout the horizon. Its depth ranges from 5 percent (high capture) to 52 percent (agentic and frontier). The senior stock declines by 7 to 14 percent by mid-2026, released at the attrition-first rate. This is the market of Section 2.1: entry-level postings down, senior postings up, experienced workers displaced in small numbers into the entry-level queue, and graduates absorbing the adjustment.

### 5.3 Phase two: reversal, 2027 onward

Table 3 reports the milestones and Figure 3 the cohort shares. Once diffusion saturates, in 2028 to 2029, junior productivity stops rising and the vendor's frontier is still advancing, at about one percentage point of task mass a year; in the baseline the firm's own frontier decays at the same time, so the aggregate frontier rises only to 0.57. Junior headcount turns up; required senior headcount continues down. In the baseline the senior stock is 25 percent below its 2022 level in 2031, 32 percent below by 2035 and halves in 2051, while required senior headcount settles about a quarter below 2022 and never halves; the junior share of headcount rises from 58 to 66 percent by 2035 and 72 percent by 2050. Absolute junior headcount does not recover its 2022 level within the horizon. The capturing firm, $\lambda = 1$, sees a faster and deeper reversal: seniors 45 percent down by 2035 and the stock halved in 2037, the junior share at 75 percent, and junior headcount 19 percent above 2022 by 2035. The reversal is the vendor's doing and the firm's own capture deepens it; the baseline is the firm that does not capture, which Section 5.7 argues is most firms in 2026.

| Scenario | Junior trough | Senior stock −25% | Senior stock −50% | Required seniors −50% | Juniors back at 2022 | Junior share 2035 | Wage bill 2035 | Tax 2035 | Tax 2050 | Pipeline 2035 |
|---|---|---|---|---|---|---|---|---|---|---|
| Baseline (α=1.6, λ=0) | 2027 Q2, -19% | 2031 | 2051 | — | not by 2052 | 66% | -21% | 1.5% | 9.1% | 0.36 |
| Capturing firm (λ=1.0) | 2027 Q1, -11% | 2029 | 2037 | 2036 | 2029 | 75% | -21% | 0.0% | 2.0% | 0.31 |
| Near-bottom capture (λ=0.25) | 2027 Q2, -16% | 2030 | 2048 | — | 2032 | 69% | -20% | 0.0% | 5.1% | 0.35 |
| Low capture (λ=0.5) | 2027 Q1, -14% | 2029 | 2044 | — | 2030 | 71% | -21% | 0.0% | 3.3% | 0.33 |
| High capture (λ=2.0) | 2026 Q4, -5% | 2028 | 2034 | 2030 | 2028 | 78% | -22% | 0.0% | 1.5% | 0.27 |
| Experiment calibration (α=1.4) | 2027 Q2, -11% | 2033 | — | — | 2031 | 66% | -15% | 0.4% | 4.7% | 0.52 |
| Frontier (α=2.0) | 2027 Q3, -35% | 2029 | 2039 | — | not by 2052 | 63% | -34% | 4.6% | 19.9% | 0.11 |
| Agentic (γ=0.5) | 2027 Q2, -38% | 2028 | 2035 | — | not by 2052 | 65% | -40% | 6.5% | 26.4% | 0.00 |
| Agentic + frontier | 2027 Q3, -52% | 2027 | 2034 | — | not by 2052 | 60% | -49% | 5.8% | 26.0% | 0.00 |
| AI-first mandate (u≥0.9) | 2027 Q2, -7% | 2030 | 2040 | — | 2029 | 72% | -18% | 8.4% | 23.0% | 0.10 |
| Long horizon (δ_F=0.10) | 2027 Q3, -17% | 2034 | — | — | not by 2052 | 63% | -17% | 0.8% | 5.7% | 0.48 |
| Fast vendor (λ_v=0.2) | 2027 Q2, -8% | 2028 | 2043 | — | 2029 | 72% | -22% | 0.2% | 7.1% | 0.34 |
| Fast renewal (μ=0.06) | falls throughout | 2034 | — | — | not by 2052 | 59% | -24% | 5.7% | 16.3% | 0.36 |
| Firm-specific share φ=0.2 | 2027 Q2, -13% | 2029 | 2045 | — | 2031 | 71% | -21% | 0.0% | 4.2% | 0.34 |
| Firm-specific share φ=0.6 | falls throughout | 2033 | — | — | not by 2052 | 61% | -23% | 4.3% | 14.9% | 0.37 |
| Frontier apprenticeship (s=0.2) | 2027 Q2, -16% | 2028 | — | — | not by 2052 | 72% | -23% | 4.1% | 4.7% | 0.59 |
| Frontier apprenticeship, agentic + frontier | 2027 Q1, -40% | 2027 | 2033 | — | not by 2052 | 71% | -42% | 5.2% | 9.6% | 0.31 |

Table 3: Milestones by calibration. Dates are the first year in which the condition holds. "Senior stock" is $n_H$ on payroll; "required seniors" is $(1 - c)/p_H$. "Junior share" is $n_L/(n_L + n_H)$ in 2035 against 58% in 2022. The wage bill includes seat costs. "Tax" is unserved tacit output as a share of task mass. The baseline is $\lambda = 0$; the capturing firm is the former baseline, $\lambda = 1$. Frontier apprenticeship raises the junior share and the pipeline because juniors take uncovered work.

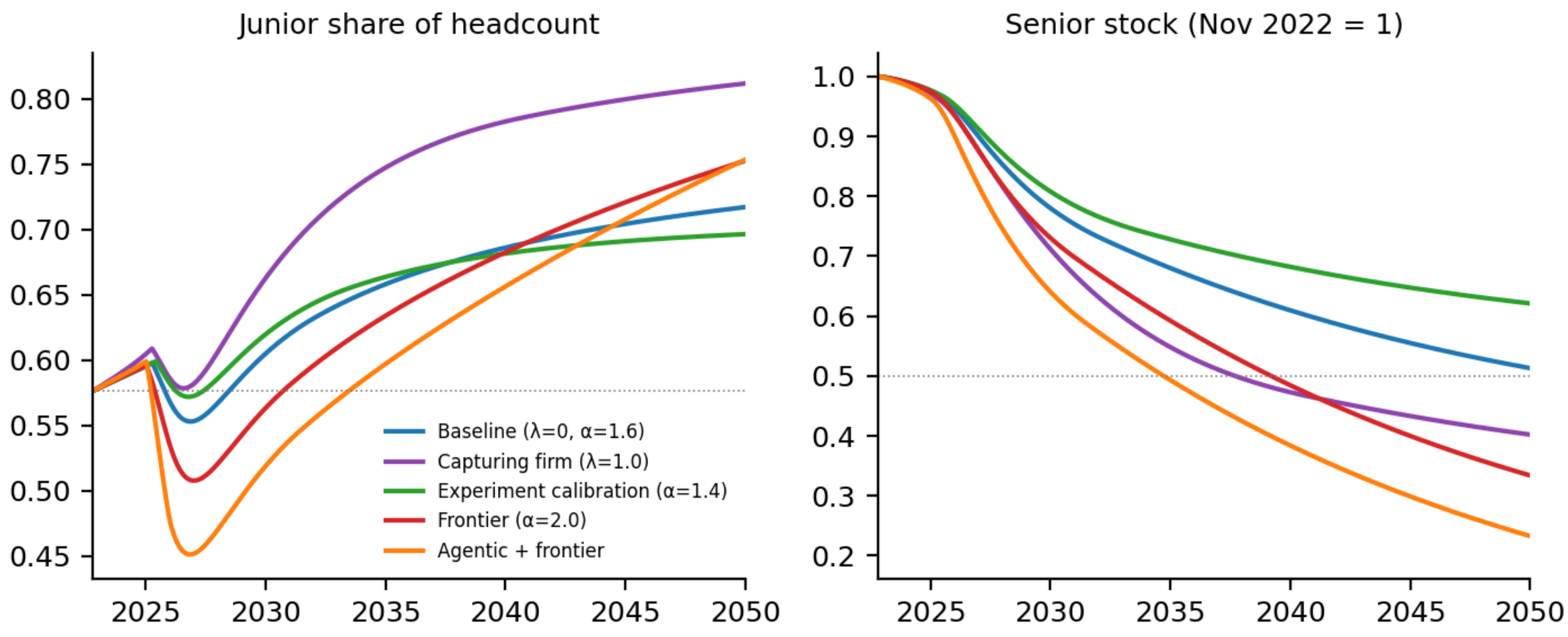

Figure 3: The cohort reversal. Left: junior share of headcount; the dotted line is the 2022 share (58%). Right: senior stock, November 2022 = 1; the dotted line marks halving.

Whether required senior headcount halves is governed by Proposition 3. In the baseline, with no capture, the aggregate frontier's steady state is $c_\infty = (1 - \varphi)c_{V,\infty} = 0.46$, with the vendor's component at 0.77 and the firm's decaying to zero, against a halving threshold $c^\dagger = 0.70$, so required senior headcount never halves; the aggregate frontier peaks at 0.57 in the mid-2030s and falls back as the firm's own documentation decays. In the capturing firm, $\lambda = 1$, the steady state is 0.73, with the firm's component at 0.67 by (6), and required senior headcount halves in 2036. The senior stock is a different quantity: it halves in 2051 in the baseline and in 2037 in the capturing firm, because the pipeline no longer replaces retirements, and Table 3 reports both. Figure 4 maps both dates over the $(\lambda, \mu)$ plane. For the stock, capture at $\lambda = 2$ moves the date to 2034; the agentic calibration, which raises senior usage $u_H$ from 0.28 to 0.52 through lower $\gamma$, moves it to 2035; combining agentic tools with the frontier calibration moves it to 2034; doubling the renewal rate removes it from the horizon. Required headcount halves only in the lower right of the plane, where capture is high and renewal slow, and nowhere at $\lambda = 0$. The date is a function of two parameters we cannot observe, and the honest statement of the result is an ordering and a condition rather than a year. The ordering, junior trough before senior decline, holds in every calibration. The condition is (6) and Proposition 3, and a reader who holds a different view of the vendor's rate or the firm's can read the date off Figure 4. We do not put a year in the abstract.

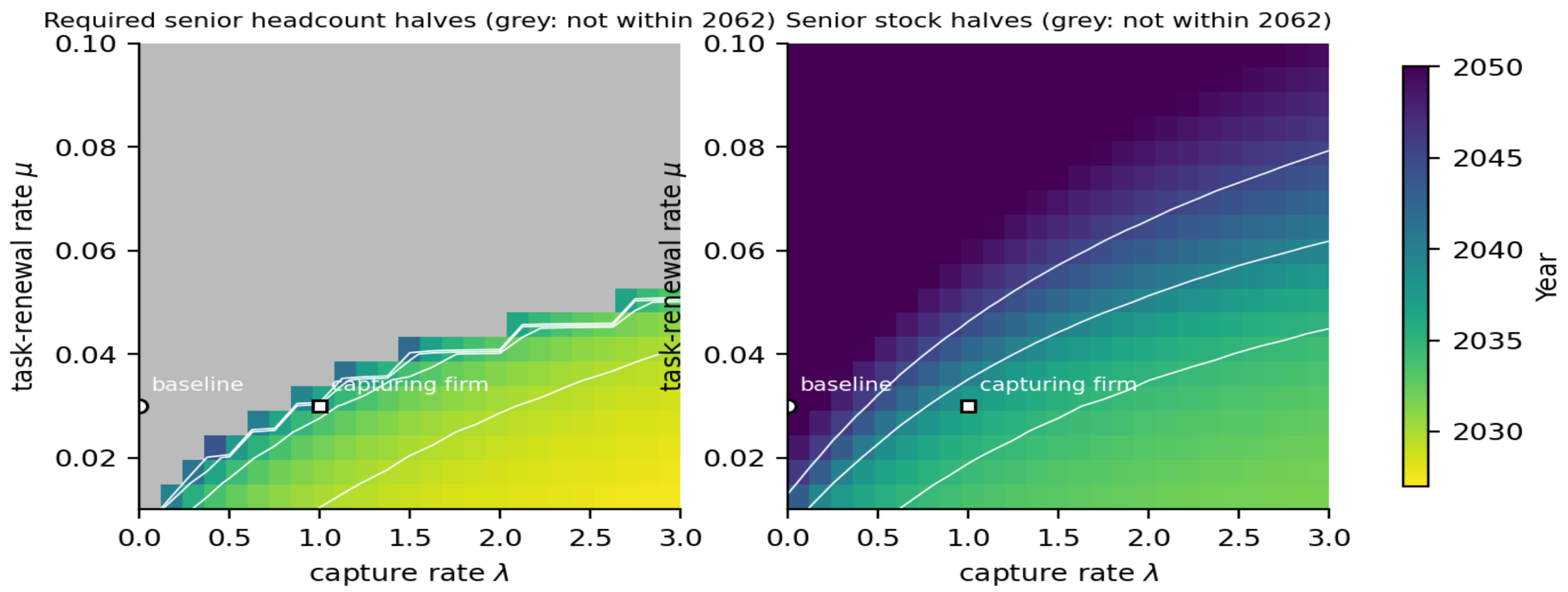

Figure 4: Year in which required senior headcount halves (left) and the senior stock halves (right), as functions of the firm's capture rate $\lambda$ and the task-renewal rate $\mu$, all other parameters at baseline ($\lambda_V = 0.10$). Grey: halving does not occur within the simulation horizon. The white contours mark 2030, 2035, 2040 and 2045. The baseline ($\lambda = 0$, $\mu = 0.03$) never halves required headcount and halves the stock in 2051; the capturing firm ($\lambda = 1$) halves required headcount in 2036 and the stock in 2037.

Two calibrations deserve separate comment because an earlier version of the baseline rejected them as descriptions of 2023 to 2026. With capture at $\lambda = 1$, an AI-first mandate that forces usage to 0.9 for both cohorts produced no junior trough, because the frontier raced ahead of junior productivity, and a vendor rate of $\lambda_v = 0.2$ did the same. At the baseline capture rate of zero both scenarios produce a shallow trough, of 7 and 8 percent, in 2027. The rejection was therefore a statement about a capturing firm under a mandate, not about mandates as such, and we withdraw it. What survives is weaker: the market of 2026 is consistent with interior usage under a short-horizon manager and a vendor frontier moving at about the baseline rate, and the depth of the observed junior decline sits between the baseline and the frontier calibration.

### 5.4 Why the junior can do it: role breadth

The reversal requires that a junior with AI serves tasks that a senior served. Section 2.1 records that roles are asking for more: 35 percent of entry-level postings require AI skills, the AI share of postings doubled in a year, and the skills requested in the average job were already changing at a third per five years before language models. In the model this is the movement of *c*. A junior in 2022 served the codified half of the distribution; a junior in 2035 serves 57 percent of it in the baseline and 70 percent in the capturing firm, and the additional points are tasks that were tacit in 2022, codified by the vendor from other firms' practice and, in the capturing firm, by this firm's seniors from their own. The junior does not do them with skill. The junior pool's mean skill falls from 0.73 to 0.39 over the same period in the baseline and to 0.11 under the agentic and frontier calibration. The junior does them with the floor, and the floor was built from the seniors, here and elsewhere, who did them with skill. Role breadth rises and role depth falls; the breadth is the tool's and the depth was the senior's. The Anthropic (2026) finding that high-tenure users delegate less and iterate more is the behavioural signature of $\beta_\tau > 1$, and the finding that newer users delegate more through directive patterns is the signature of $\beta_c < 1$ on the tasks they are given.

This is also the answer to the objection that seniors' tacit knowledge protects them. It does, as long as the tasks that require it remain tacit, and on the vendor-coverable set that is not the firm's decision. On the firm-specific set Proposition 3 says that the seniors' own usage moves the tasks across the frontier, and that a shorter evaluation horizon moves them faster. The protection is real, the firm controls it only on its own residual, and there it is consumed by the act that produces the productivity gain.

Proposition 5 gives the sharper form, and the calibration gives it a size. If senior formation requires practice on uncovered work, and juniors in 2035 spend their time on codified work at usage 0.57, the pipeline index is 0.36 in the baseline: nearly two-thirds of the formation rate is gone. Almost all of that loss is delegation. At $u_L = 0$ the index would stand at 0.98 in the baseline and at 0.84 in the capturing firm, because the frontier moves little without capture and the juniors the codified mass requires offset the ground it removes; the shrinking of the practice ground is the mechanism of the proposition, and at these frontier rates it is the smaller channel. The frontier apprenticeship scenario in Table 3 assigns a fifth of uncovered work to juniors unaided: the pipeline index rises to 0.59, the junior pool's skill in 2035 is 0.48 rather than 0.39, the rebuild time falls from 9.0 to 8.3 years, and the firm pays for it in a visible quality shortfall of 4 percent of task mass every year. Under the agentic and frontier calibration the same lever moves the pipeline from zero to 0.31. That is the only mechanism in the model by which a firm whose codified core belongs to the tool continues to produce seniors, and it costs current output, which is why a manager evaluated on current output will not choose it.

### 5.5 Pricing the transfer

The companion workbook prices what one senior expert spent to reach the capability that a tool now installs in a junior: statutory tuition and materials over four study years, entry salary forgone while studying, self-funded courses and memberships, and five discretionary learning hours a week over twenty working years valued at the average hourly wage, each restated in 2026 euros at a three percent real rate. The total is €433,000 per expert, of which €401,000, 93 percent, was borne by the worker and €32,000 by employers. The workbook measures from the receiving end, valuing what the junior received at what it would have cost to build the old way, and it is explicit that the figure is a measure of scale and not a bill.

The payer split is the line that survives scrutiny. The ratio of capability received to seat price paid, which the reader will compute from the table, does not: every capital good that reproduces at near-zero marginal cost shows the same ratio, and a textbook contains centuries for the price of a paperback. What the textbook analogy does not reach is that the textbook's author was paid for the textbook. Here the person who financed the capability, out of forgone earnings and unpaid hours, was paid a rent for it while it was scarce, and the point of the reproduction is that the rent ends earlier than the career over which it was expected to run. Sunk to the firm, the build cost was financed by the worker against a premium that the reversal ends early. That is a distributional fact, and the paper does not make it an efficiency claim.

Table 4 sets the transfer out for a reference team of ten seniors and fourteen juniors in 2022, scaled from the model's unit task demand, for the baseline and the capturing firm in 2035, the frontier and agentic calibrations at the years in which each becomes fully visible, and frontier apprenticeship in the baseline. The senior stock declines for two reasons that the ledger separates: surplus seniors released, and retirements the pipeline no longer replaces. Only the first is a released senior with something to lose, and the loss is the present value of the seniority premium forgone over fifteen remaining years at a re-employment salary of €55,000, an assumption informed by the Indeed finding that workers with more than ten years' experience are now the largest group applying to entry-level roles. The build cost is not added to that loss: it is sunk, and the worker earned a return on it for as long as the premium was paid. It enters the ledger as what it cost to build what the tool now substitutes, a stock valued once at build cost times the senior-equivalents no longer required, set against the seat cost the team pays in the year. The organisation's saving is the annual reduction in the loaded wage bill against 2022, seats included. The tax is the unserved tacit output valued at the 2022 cost of a unit of task mass, which is a lower bound because tacit tasks are the higher-value tasks; alongside it we report the frontier capacity lost and the rebuild time, which have no price in the account.

| | Baseline, 2035 | Capturing firm, 2035 | Frontier, 2040 | Agentic + frontier, 2050 | Baseline + apprenticeship, 2035 |
|---|---|---|---|---|---|
| Senior stock (2022: 10, or 8 under apprenticeship) | 6.8 | 5.5 | 4.8 | 2.3 | 6.2 |
| Required seniors (of 10) | 7.2 | 5.0 | 7.0 | 6.1 | 5.8 |
| of the decline: released (surplus) | 1.3 | 2.9 | 1.3 | 1.9 | 3.6 |
| of the decline: retired and not replaced | 1.9 | 1.7 | 3.8 | 5.7 | 0.3 |
| Loss per released senior (PV of premium forgone, 15 yrs) | €358k | €358k | €358k | €358k | €358k |
| Aggregate expert loss | €0.48 m | €1.02 m | €0.48 m | €0.69 m | €1.27 m |
| Senior-equivalents the tool substitutes | 2.8 | 5.0 | 3.0 | 3.9 | 4.2 |
| Capability substituted, at build cost (stock) | €1.19 m | €2.17 m | €1.32 m | €1.69 m | €1.82 m |
| of which financed by the workers who built it | 93% | 93% | 93% | 93% | 93% |
| Juniors in the year (of 13.6) | 13.1 | 16.2 | 10.4 | 7.1 | 15.9 |
| Seat cost in the year, whole team | €11.9k | €13.0k | €9.2k | €5.7k | €13.3k |
| Loaded wage bill 2022 → year | €1.74 m → €1.38 m | €1.74 m → €1.38 m | €1.74 m → €1.03 m | €1.74 m → €0.60 m | €1.87 m → €1.44 m |
| Annual organisational saving | €367k (21%) | €366k (21%) | €712k (41%) | €1147k (66%) | €429k (23%) |
| Unserved tacit output, share of task mass | 1.5% | 0.0% | 10.5% | 26.0% | 4.1% |
| of which firm-specific | 0.0% | 0.0% | 0.0% | 12.8% | 0.0% |
| valued at 2022 cost (lower bound) | €26k | €0k | €183k | €454k | €76k |
| Net position in the output account | €+341k | €+366k | €+528k | €+694k | €+353k |
| Frontier capacity lost | 41% | 67% | 58% | 79% | 47% |
| Pipeline index | 0.36 | 0.31 | 0.10 | 0.00 | 0.59 |
| Junior pool skill S_L | 0.39 | 0.39 | 0.19 | 0.07 | 0.48 |
| Years to rebuild one senior (4.9 in 2022) | 9.0 | 9.0 | 10.5 | 11.2 | 8.3 |

Table 4: The transfer priced three ways, for a team of ten seniors in November 2022 (eight seniors and three apprentices under frontier apprenticeship, at the same task demand). Build cost from the companion workbook (€433,377 total, €401,133 worker-borne); premium forgone = (€85,000 − €55,000) × 11.94 (fifteen-year annuity at 3% real). Released seniors are the cumulative release of surplus; the rest of the decline is retirement the pipeline did not replace. The companion workbook's Ledger sheet has not yet been rebuilt to this layout.

Three things are visible in the table. The saving is large, immediate and grows with the reversal: a fifth of the wage bill in the baseline by 2035, two-thirds under the agentic and frontier calibration by 2050. The measurable tax is small and late in the baseline, 1.5 percent of task mass in 2035, worth seven percent of the saving, and none of it on the firm's own residual; it is neither small nor late under the frontier and agentic calibrations, where it reaches 10 and 26 percent of task mass and a quarter to two-fifths of the saving. The unmeasured tax is large and early: two-fifths of frontier capacity is gone by 2035 in the baseline, two-thirds in the capturing firm, and the time to regenerate a senior has roughly doubled. A manager with a three-year horizon sees the first line, is charged for the second only when it arrives, and cannot see the third. Discounting the whole path at $\delta_F = 0.33$, the present value of the saving per unit of task demand is €14,000 in the baseline and €36,000 under the agentic and frontier calibration; the present value of the measurable tax is below €1,000 in both at that rate, and reaches €36,000 at the worker's rate and €190,000 at a two percent social rate under the agentic and frontier calibration. The wedge between the manager's and the worker's valuation of the same path is the wedge of Proposition 5 in the original paper, now on the firm's own books. The apprenticeship column shows the price of keeping the pipeline: a visible quality cost of 4 percent of task mass, €76,000 a year, for a pipeline index of 0.59 rather than 0.36 and a senior stock that stays within two percent of requirement through 2050.

### 5.6 The adaptability tax

Figure 5 shows the three measures of Section 3.5. In the baseline, junior usage stays interior at 0.57 and the pipeline runs at 0.36 of its 2022 rate; that is below what the uncovered set requires, so the senior stock falls below requirement in 2032. The output tax is zero until then, 1.5 percent of task mass by 2035 and 9 percent by 2050, almost all of it on vendor-coverable work, and frontier capacity falls to 47 percent of its 2022 level by 2050. In the capturing firm the frontier is higher, requirement lower and the stock stays above it until 2041; its tax is 2 percent by 2050 and its frontier capacity 22 percent. Under the frontier and agentic calibrations junior usage reaches or approaches one, the pipeline falls to 0.10 and to zero, and the senior stock, no longer replenished, decays at the retirement rate; the output tax appears in 2031 and rises to between 20 and 26 percent of task mass by 2050, of which 8 to 13 points fall on the firm's own residual, and frontier capacity falls to between 21 and 30 percent of its 2022 level. This is Proposition 4 with its pipeline condition: the tax is conditional on the pipeline, not on the frontier, and a deployment that stays interior still pays it once the pipeline it leaves open is too thin to replace retirement. The seniors the firm keeps are spent on its own residual first, which is why the firm-specific component stays near zero in the baseline through 2050 and appears only in the calibrations that close the pipeline. The firm-level result is that the adaptability tax arrives through the pipeline and is a capability fact first: frontier capacity falls a decade before the first unit of output is lost, and the three panels of Figure 5 are the place to see it.

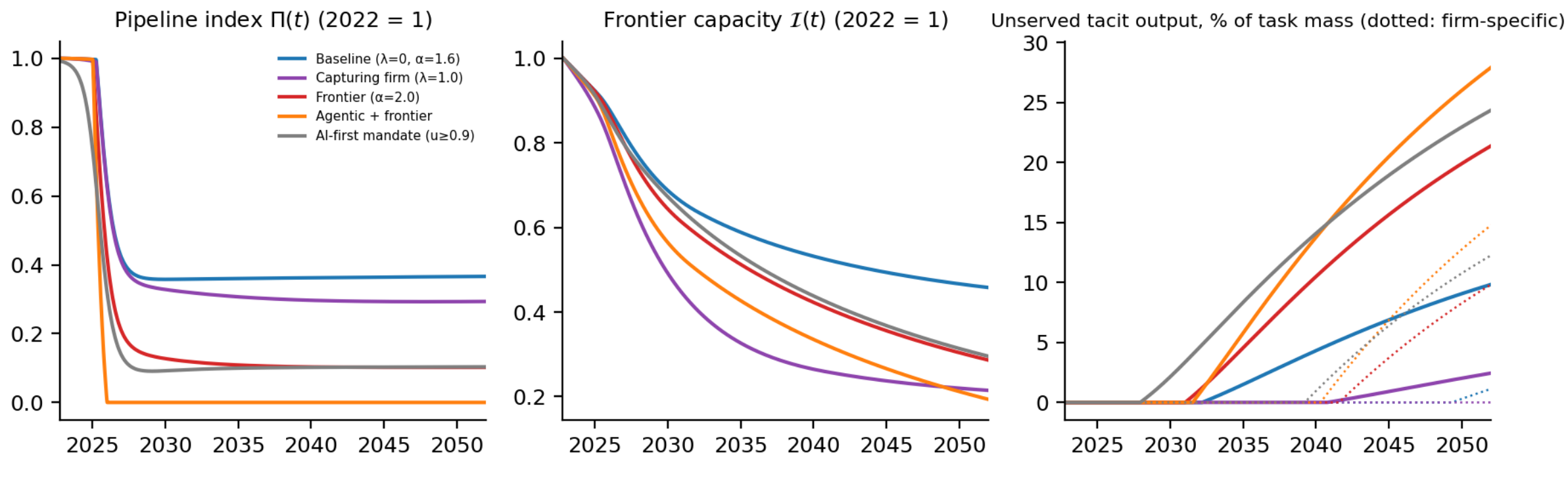

Figure 5: The adaptability tax. Left: pipeline index over uncovered work, equation (7). Centre: frontier capacity $\mathcal{I}_t = n_H(1 - c)$, November 2022 = 1. Right: unserved tacit output, equation (9), as a percentage of task mass; dotted lines are the firm-specific component, which stays near zero in the baseline and the capturing firm and appears after 2040 in the calibrations that close the pipeline. The AI-first mandate holds usage at 0.9 for both cohorts from adoption.

The renewal shock in Figure 6 asks what happens when the environment changes after the seniors have gone. In 2040 we reset both frontiers by 25 points, the model's representation of a regulatory regime change, a new product category or a technology shift that returns a quarter of the firm's work to the tacit set. Every firm is hit, because no firm holds seniors for tasks that were codified. The difference is in recovery, and the free-rider correction splits it in two. The vendor-coverable frontier recovers in every scenario, from 0.49 to 0.71 by 2052 against a pre-shock 0.74, because the vendor's rate does not depend on this firm's seniors; the output tax on that work falls with it. In the capturing firm the total tax falls from a peak of 15 percent of task mass to 5 percent by 2052 and the firm-specific component stays at zero. The firm-specific frontier is informative only where the firm captures: in the capturing firm it stood at 0.69 before the shock and at 0.59 twelve years after, a recovery of 15 of the 25 points lost against the vendor's 22. In the baseline there is nothing to recover, because without capture $c_f$ had already decayed to 0.30 by 2040, and the shock takes it to 0.05, where it stays; the tax after the shock is 16 percent of task mass in 2052, of which 6 points are firm-specific, and under the agentic and frontier calibration 37 and 21 percent. The firm has too few seniors to re-encode its own residual faster than renewal erodes it, and the pipeline that would produce more is at a third of its former rate or at zero. What the firm has lost is the ability to codify its own practice again, and the measure of that is $c_f$.

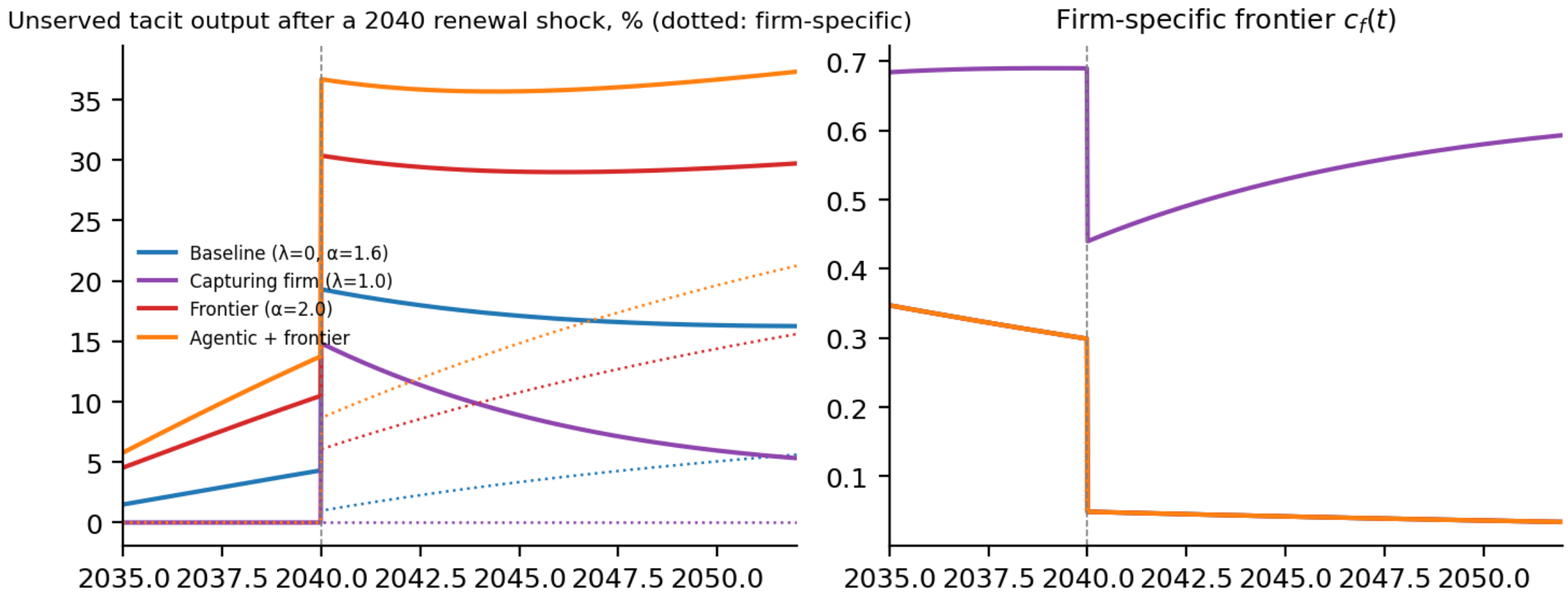

Figure 6: A renewal shock in 2040. Both frontiers are reset by 25 points of task mass. Left: unserved tacit output; dotted lines are the firm-specific component. Right: the firm-specific frontier $c_f$. The vendor frontier recovers most of the lost ground in every scenario; the firm-specific frontier recovers part of it in the capturing firm and none of it where the firm does not capture.

The shock experiment is the model's operational definition of adaptability and, by extension, of innovation. A firm innovates in this model when it serves tasks the tool has not seen, and it can do so only with workers above $S_{min}$. Frontier capacity $\mathcal{I}_t$ is that ability measured before the shock arrives, and the recovery of $c_f$ is that ability measured after. A firm in which AI works with juniors and few seniors remain has an $\mathcal{I}_t$ near zero, which means that it is fully productive on the task distribution as codified, that it will be fully productive again on vendor-coverable work when the vendor catches up, and that it can no longer move on its own work when its own work moves. Creativity in the ordinary sense, the recombination of judgment across tasks that have not been posed before, is the uncovered set by definition, and it is served by the cohort the reversal removes and formed, by Proposition 5, only on the tasks the reversal has made scarce.

### 5.7 Locating a firm on the parameter axes

The results above are reported as regions because two parameters are not observable in the data available to us. A region is informative only to the extent that a reader can place a real firm on the axis it is drawn over, so this section says what we would measure, where we think firms stood in 2026, and what would change our minds. McEntire (2026), reviewing the same literature, identifies the entry-level slowdown as consistent with firms automating the routine tasks that traditionally served as apprenticeship pathways and treats it as a segment-level friction; the model here is an attempt to state that friction as a mechanism and to price it.[3]

The firm's capture rate λ is a property of deployment design, and deployment design is observable from the inside without any new instrument. Three features carry most of it: whether the firm retains and reuses the corrections its seniors make to model output, whether approvals, overrides and rejections are logged in a form that can be learned from, and whether agents act under senior sign-off in a way that records what the senior changed. A firm with none of these is near the bottom of the range on its own residual whatever its licence spend; a firm with all three is near the top. Our prior for 2026 is that most firms sit close to the bottom, because correction traces are discarded by default in every deployment we have seen, and what would move us is evidence that routine enterprise deployments retain and reuse them without anyone deciding to.

[3] The proposal to treat deployment features as observable proxies for the capture rate is due to Jeremy McEntire, in correspondence, September 2026, together with the objection that a region without a way of locating a firm within it is a hedge with a coordinate system attached.

The firm-specific share φ is harder, and the natural proxy does not survive contact with the model. Measuring the share of senior time spent outside vendor tools conflates φ with senior usage, because capture in (5) runs precisely through seniors working on firm-specific tacit tasks inside the tool. Two measurements do not have that defect: the performance gap between the base model and a firm-tuned model on the firm's own tacit tasks, which is an estimate of what remains uncodified for that firm; and the share of senior corrections that turn on firm-specific context rather than general capability, which can be read from a sample of logged corrections where any exist.

The identification burden differs between the two, and at the baseline it is no longer one-sided. Table 5 reports the model at 2035 across the plausible range of each parameter with the other held at baseline. The pipeline index is close to flat in both: it moves by five points over the range of φ and by nine over the range of λ. The frontier and the tax are not. Without capture the firm-specific share is the part of the task mass that can never be codified, so over the range of φ the frontier in 2035 moves from 0.68 to 0.42, the 2035 tax from zero to 8 percent of task mass, and above φ = 0.5 junior headcount falls throughout instead of recovering; over the range of λ the frontier moves from 0.57 to 0.75 and the tax from 1.5 percent to zero. A reader who disagrees with us about either parameter is reading a different timetable for the tax and the same timetable for the pipeline, which is the right place for the disagreement to sit.

| Parameter | value | Pipeline 2035 | Stock/req 2035 | Frontier 2035 | Junior trough | Tax 2035 | Junior share 2035 |
|---|---|---|---|---|---|---|---|
| φ | 0.1 | 0.321 | 1.061 | 0.677 | 2027.2 | 0.0% | 73% |
| φ | 0.2 | 0.338 | 1.018 | 0.641 | 2027.3 | 0.0% | 71% |
| φ | 0.4 | 0.360 | 0.939 | 0.567 | 2027.4 | 1.5% | 66% |
| φ | 0.6 | 0.367 | 0.850 | 0.494 | none | 4.3% | 61% |
| φ | 0.8 | 0.358 | 0.767 | 0.421 | none | 7.6% | 57% |
| λ | 0.0 | 0.360 | 0.939 | 0.567 | 2027.4 | 1.5% | 66% |
| λ | 0.25 | 0.347 | 1.001 | 0.616 | 2027.3 | 0.0% | 69% |
| λ | 0.5 | 0.333 | 1.040 | 0.653 | 2027.2 | 0.0% | 71% |
| λ | 1.0 | 0.307 | 1.099 | 0.702 | 2027.1 | 0.0% | 75% |
| λ | 2.0 | 0.274 | 1.160 | 0.752 | 2027.0 | 0.0% | 78% |

Table 5: Sensitivity of the 2035 outcomes to the two unobserved parameters, each varied over its plausible range with all others at the baseline of Table 2. The pipeline index is close to flat in both; the frontier, the senior stock relative to requirement and the tax depend on both, so the observables of this section matter for the timing of the tax and not for the pipeline.

One further result belongs here, because it bears on how much of the argument depends on capture at all. The baseline is the free-riding firm, λ = 0, which contributes nothing to any model, and the ordering holds in it: the junior trough arrives in 2027 and the senior decline follows it. The ordering does not require the transfer to happen; it requires only that the vendor's frontier keeps moving, which is the market-level form of Assumption 1 and the one supported by the Stack Overflow series of Section 2.2.

## 6 Discussion

### 6.1 Objections, in the order a reader will raise them

The results above were reached by stress-testing a stronger thesis until it broke, and the order in which its parts broke is the order in which a reader will object. We keep the trail in, because the survivor is credible only against what it survived.

*The reallocation objection.* This is the oldest story in labour economics: every general-purpose technology embodies skill in capital and moves the work to cheaper labour, and nobody mourns the in-house typesetter. Most of an earlier draft's headline collapses into this. The reversal is capital-labour substitution and will be chosen; the expert's loss is a rent the market has stopped paying; even the tax is, in part, ordinary depreciation that the firm manages by buying the next release. What the reallocation story cannot reach is the maintenance input. In every previous embodiment the capital good was built and maintained by a different population of experts who kept practising: CAD was written by engineers who still drew. Here the capital good is built from the exercise of the skill by the population that is ceasing to exercise it, and the vendor does not practise it either. A lathe does not need lathe operators to remain a lathe. The codified stock needs practising experts to remain current against a changing task distribution, and that is the first statement of Assumption 1.

*The free-rider objection.* The source is not the firm's seniors; the model is trained on the world, and one firm's ten seniors are a rounding error against the vendor's corpus. This objection is correct and the paper now incorporates it: the vendor's frontier in (5) has no senior in its source term, and the firm's exposure is only its firm-specific residual. Two things survive. The first is that the objection is the mechanism one altitude up: if every firm rationally free-rides, the practising population for the shared model is the vendor's contractors and whoever still does the work unaided, which is the collapse state of Acemoglu, Kong and Ozdaglar (2026) and the Stack Overflow series of Section 2.2. The second is that the residual, though small, is the part no vendor can codify and no vendor can regenerate, and Section 5.6 reports that it is the part that recovers least from a shock. The synthetic-practice version of the objection, that the model can generate its own practice, holds where verification is cheap and fails on judgment tasks with slow, ambiguous feedback; and where human signal is still needed, the vendor's contractor is exactly the practising expert the argument requires.

*The new-senior objection.* Juniors become seniors anyway, of a different kind: they learn to specify, decompose, verify and orchestrate, and that is the $S_{\min}$ of 2040. Part of this holds and the model now reflects it: the pipeline index of the earlier draft counted practice on the junior's own codified tasks, which builds fluency the tool already supplies. What refuses is the loop through which a junior learns that the tool was wrong. Either a senior catches it, which requires seniors; or consequences arrive, slowly and silently; or the junior does the task unaided and compares. On tacit tasks, where by definition the tool's output cannot be trusted, competence is built by doing them, so the new kind of senior can only be formed on uncovered work. That is Proposition 5, and it is why the ladder inverts rather than merely lengthens.

*The aided-practice objection.* The pipeline index counts unaided practice, and a junior who works through the tool and carries the consequence of a bad output may be building precisely the competence the job now requires, which is the tool's failure map. Equation (2) does not count that, and the omission is a modelling choice rather than a finding. Two things survive it. The uncovered set shrinks whatever the mode of practice, so a firm can form aided judgment at scale and still lose the capacity priced in Section 5.6; and the consequence has to reach the learner, which is a matter of allocation. Where agents run under senior sign-off the miss is attributed to the senior, and the loop forms the senior. That is the lever of Proposition 5 in a different currency, incidents rather than a visible quality shortfall. Whether aided practice with attributed consequence substitutes for unaided practice on the firm-specific residual is the sharpest question this paper leaves open, and we do not settle it.[4]

*The date objection.* The senior halving date depends on two parameters the paper cannot observe, so it is a parameter choice with a year attached. This is correct, and the year has left the abstract. What survives is the ordering, which holds in every calibration and can be tested against the payroll series by 2028, and the condition, which lets a reader with a different view of the vendor's or the firm's rate read the date off Figure 4. The rejection of the mandate and fast-vendor scenarios as descriptions of 2023 to 2026, which an earlier version listed here, does not survive the move of the baseline to $\lambda = 0$ and is withdrawn in Section 5.3.

*The equivalence objection.* A reader coming to this from the demand-externality literature will ask whether capability erosion adds anything to a model that already carries a reabsorption rate. Takuma Kimura, whose extension of that literature endogenises the reabsorption rate, has pointed out in correspondence that at a symmetric configuration it may not. Under congestion the reabsorption rate depends on the sector's average automation, so a single firm moves that rate only by its share of the sector while the deterioration reaches every displaced worker; under erosion the rate responds fully to the firm's own automation while the deterioration reaches only that firm's displaced workers, who are the same fraction of the total. The two offset exactly, leaving the marginal demand loss, the Nash and cooperative automation rates and the corrective tax of Hemenway Falk and Tsoukalas (2026) unchanged. Three things escape the equivalence. It holds only at symmetry, and once automation rates differ across firms erosion predicts a strictly larger aggregate loss, and it predicts a dispersion in re-employment outcomes that congestion does not, which is the sixth prediction of Section 6.5. It removes the best-response dependence that congestion creates between firms, since a firm's automation no longer degrades the reabsorption of anyone else's displaced workers. The demand externality itself survives, through the spending those workers no longer do, which is why the corrective tax is unchanged; what becomes private is the capability loss, which is borne by the firm that caused it and therefore behaves at firm level as the agency problem of Section 6.2 rather than as a second externality. And accumulated exposure, how long a firm has automated rather than only how far, is a history a static specification cannot carry, which is the reason the model here is dynamic.[5]

[4] This objection, and the observation that answerability is the same lever as the assignment share, are due to Jeremy McEntire, in correspondence, September 2026.

*The ledger objection.* A €600 seat containing €433,000 of build cost is the ordinary economics of reproduction at near-zero marginal cost; a textbook shows the same ratio. This is correct and the ratio has left the prose. What survives is the payer split, for the reason given in Section 5.5: the textbook's author was paid for the textbook, and the person who financed 93 percent of this capability was paid a rent for it only while it was scarce. The build cost itself is sunk and is no longer counted in the expert's loss; the loss is the premium forgone.

### 6.2 The organisational augmentation trap

Caosun and Aral define the augmentation trap as the case in which the worker's lifetime welfare under the equilibrium usage policy falls below welfare absent AI, and show that it arises when the decision-maker's horizon is shorter than the worker's or when the firm ignores the worker's private return to skill. The cohort model produces the same structure one level up. The decision-maker is the manager with $\delta_F$; the party bearing the cost is the organisation over a horizon set by its senior careers and its exposure to renewal; and the cost is the adaptability tax. The manager is not mistaken. Every line of Table 4 that a manager sees is positive, and the present value of the saving at the manager's rate exceeds the present value of the measurable tax by more than an order of magnitude in every calibration without apprenticeship. The trap is chosen because the account in which it would appear does not exist. The same mechanism now has evidence at three altitudes: the individual decision, where a fluent answer lowers the threshold at which a worker judges they know enough to answer (Marcoccia, Quattrociocchi and Capraro, 2026); the firm, which is this paper; and the economy, where the effort that produces general knowledge is substituted away (Acemoglu, Kong and Ozdaglar, 2026).

The original paper's diagnosis of what output metrics can and cannot distinguish carries over unchanged. AI raises short-run productivity in every adoption region, so output cannot distinguish a deployment in Region IV from one in Region II; here, wage-bill savings rise in every calibration, so the wage bill cannot distinguish a reversal with an intact pipeline from one without. The measures that can are the three of Section 3.5. The pipeline index is observable from task allocation and unassisted assessment: it is junior practice on uncovered work. Frontier capacity is observable from task allocation: it is senior time on work the tool does not handle. The rebuild time is a calculation from the pool's assessed skill. None of these is in a management account today, and the original paper's argument for periodic unassisted assessment applies with more force to a cohort than to a worker, because a cohort that has stopped learning is not visible in any individual's output.

### 6.3 What moves the outcome

Five parameters set the timing and the depth of the reversal and whether the pipeline survives it, and each corresponds to a lever. The vendor's rate $\lambda_v$ is not the firm's to set; it is set by the market-level form of Assumption 1, and it determines most of the reversal's timing. The firm's capture rate $\lambda$ is the analogue of workflow design: it is set by whether the firm logs and learns from senior corrections and lets agents act under senior approval, and it governs the firm's own residual and nothing else; a firm that wants to keep its residual tacit lowers it, at the cost of a tool that improves more slowly on that firm's own work. The evaluation horizon $\delta_F$ sets senior usage and thereby the firm-level source term of (5). The task-renewal rate $\mu$ is largely exogenous, varies by sector, and the sectors in which regulation, products and technology renew fastest are the sectors in which the reversal is slowest and the firm-specific tax, when it comes, arrives soonest. The position of the junior deployment relative to the automation boundary $C_1$ determines whether the pipeline survives, and it is set by $\alpha$ relative to $\gamma$, by mandates, and by the design of the deployment itself. The last of these is the one the firm holds. Bastani et al. (2025) put the same model in front of comparable students under two designs and obtained a 17 percent decline in unassisted performance under one and no decline under the other, which is the difference between a junior deployment inside the adoption region and one in the automation region, at fixed model capability. The AI-first mandate in Table 3 is the clearest case of the choice going the other way: it produces a large junior share, a small wage saving, a pipeline index of 0.10, and a positive tax from 2028, the earliest of any scenario, which is a worse position for the organisation on every measure than the baseline, chosen for the sake of a compositional target.

[5] We are grateful to Takuma Kimura for this argument, in correspondence, September 2026, and for the observation that the reabsorption rate of Hemenway Falk and Tsoukalas (2026), defined as the share of a lost wage recovered through new employment or transfers, already accommodates re-employment at a lower wage without redefinition.

**6.4 Responses**

The responses in Section 3.2 of the original paper, training, evaluation period and workflow design, remain the responses here, with two additions specific to the cohort structure. The first is frontier apprenticeship, Proposition 5's lever: a share of uncovered work assigned to juniors unaided, at a quality cost that appears in current output. In the baseline it lifts the pipeline index from 0.36 to 0.59 for a visible cost of 4 percent of task mass a year, and keeps the senior stock within two percent of requirement through 2050; under the agentic and frontier calibration it lifts it from zero to 0.31. It is the only mechanism in the model by which a firm whose codified core belongs to the tool continues to produce seniors, and it is the mechanism a manager evaluated on current output will not choose, because its cost is the one line in Table 4 that the output account does see. The second is deliberate frontier management on the residual: because the firm-level source term of (5) is senior usage on firm-specific tacit tasks, a firm can choose which of those tasks its seniors exercise through the tool and which they exercise without it, and can thereby choose which of its own capabilities it still holds in fifteen years. The original paper notes that Zuboff (1988) found organisations choosing to automate rather than inform when the same system could do either; the choice is now about the firm's own residual, and it is the only part of the frontier the firm controls.

The guardrailed arm of Bastani et al. (2025) is the closest existing instance of this lever. The restriction that withheld solutions and required the student's own working preserved unassisted performance while more than doubling assisted performance, which is the shape frontier apprenticeship takes in a deployment rather than in a task allocation. Whether the same restriction preserves formation on tacit work with slow and ambiguous feedback is untested.

A third response follows from the same evidence and lies outside the model as written. A deployment that reveals the tool's answer only after the learner has attempted the task separates practice usage from delivered usage: the attempt runs unaided while the output the firm ships is the tool's. The model here carries a single usage variable governing both production in (1) and learning in (2), so it cannot represent that separation, and the quality cost that makes frontier apprenticeship expensive in Table 4 is, on the band of work immediately above the coverage line, partly an artefact of that restriction. On the firm-specific residual the restriction binds for a substantive reason rather than a formal one: there the tool's output cannot be relied upon by construction, so there is less to reveal and the quality cost stands. Separating practice usage from delivered usage is a change to the model rather than a paragraph within it, and it is the first extension we would make.[6]

A fourth response goes further, and the model bounds how far. Formation in this paper runs on consequences the practitioner owns and waits for, which is why the loss of practice looks like a loss of capacity. It need not be the practitioner's own consequences. Historians are formed on decisions that resolved centuries before they were born, and what their training supplies is not access to the record but a method for reading it: reconstruct the decision as it stood before the outcome was known, resist hindsight, compare across cases. Software has an unusually complete corpus of exactly this kind, in postmortems, incident reviews and migration write-ups, and almost nobody is taught to read it that way. If the method can be taught, then the binding constraint on formation is the distribution of unflinching accounts rather than the capacity to produce experts, which is a more tractable problem and a different paper. The model bounds the claim rather than settling it. A public corpus exists for the vendor-coverable share of tacit work and does not exist for the firm-specific residual: there is no published record of this firm's clients, its legacy systems or its regulator's habits, and $\varphi$ is the parameter that separates the two. Two further cautions apply on the share where a corpus does exist. Postmortems are written about failures that were survivable and publishable, which selects against precisely the out-of-distribution case Section 5.6 prices. And the method of reading them is itself taught, which returns part of the regress it was meant to escape.[7]

**6.5 Testable predictions**

The model adds five predictions to those of the original paper, and the equivalence argument of Section 6.1 adds a sixth, which concerns a re-employment market the model does not contain.

---

[6] The structural point, that a look-first deployment decouples practice usage from delivered usage and that a single usage variable cannot represent it, is due to Jeremy McEntire, in correspondence, September 2026.

[7] The argument that the scarcity is one of distribution rather than of capacity, and the analogy to historical method, are due to Jeremy McEntire, in correspondence, September 2026. The bound on it is ours.

First, *the junior trough precedes the senior decline and its timing tracks adoption diffusion, not model capability.* Across sectors, the date at which early-career employment in exposed occupations stops falling should coincide with saturation of workplace adoption and should be similar across sectors with different α; the depth of the trough should scale with the junior productivity gain measured in that sector's experiments. The payroll series that produced the 2026 gap can test this directly by 2028.

Second, *senior displacement is increasing in the firm's capture rate, within occupations.* Firms that fine-tune on expert output, deploy agents under senior approval and log review decisions should show senior headcount falling earlier and further than firms using the same models without capture. Across occupations the vendor's rate dominates and the prediction is weaker.

Third, *the reversal is compositional before it is absolute.* The junior share of headcount should rise in every exposed occupation after the trough; absolute junior headcount should recover only where the frontier advances faster than junior productivity, which the model associates with lower-α, lower-γ deployments.

Fourth, *the entry task set in exposed occupations is getting harder.* If senior formation has moved to the frontier, the task content of entry-level postings in exposed occupations should be rising in seniority over 2023 to 2026, and the share of entry-level roles that consist of codified work should be falling. The 35 percent of entry-level postings requiring AI skills is the wrong measure of this; the right measure is the seniority of the tasks listed, which is a Lightcast pull. This is the prediction that would confirm or kill Proposition 5 with data that already exist.

Fifth, *adaptability falls before output does, and firm-specific capability falls last and recovers never.* Firms whose junior deployment lies in the automation region should show, in order, a collapse of unassisted junior performance, a decline in senior time on unstructured work, and, only after a renewal shock, an output loss, with the loss on common work recovering as the vendor catches up and the loss on firm-specific work persisting. A panel of firms observed through a regulatory regime change, with pre-shock measures of junior unassisted performance and senior time allocation, would test the ordering. The RLHF and expert-annotation labour market is, on this argument, the last place the common skills are exercised with stakes, and it should be studied as such.

Sixth, *within an occupation, the seniors released first are those whose employers captured most.* Erosion and congestion in reabsorption both predict that a displaced worker recovers less of a lost wage as automation spreads, and they differ in what that fall is a function of. Congestion makes it a property of the sector, so all displaced workers in a labour market face the same deterioration whatever their employer did. Erosion makes it a property of the firm: a worker released from an employer that logged and reused senior corrections for five years has lost more of what the next employer would pay for than a worker released from one that did not. The prediction is therefore about dispersion rather than level. Within a single occupation and a single displacement wave, re-employment wages should vary systematically with the prior employer's capture rate, proxied by the deployment features of Section 5.7, and they should not vary that way if congestion is the whole story. This is the observable that separates the two mechanisms of the equivalence objection in Section 6.1.

## 7 Limitations

The senior threshold is a scalar, and seniority is not. $S_{\min}$ is crossed once in this model, after which a senior is a senior. Practitioners describe something with more structure: serving the firm-specific residual, advancing the frontier on work nobody has done before, and forming the next person are three capabilities, acquired roughly in that order, and the third is exercised by supervising how others exercise the first two. The model cannot see the difference, and the omission bites at exactly the point the paper is about. Release in (8) removes surplus seniors at a constant rate without asking which capability each one carries, so two firms shedding the same number of seniors end up with different organisations. A last-in-first-out rule sheds people who have only the first capability and preserves the third; attrition-first sheds the most marketable, who are disproportionately those who have the second and the third, and it is attrition-first that the calibrations here assume. The prediction that follows is testable and, as far as we know, untested: among firms that reduce senior headcount by comparable amounts, those that do so through voluntary attrition should lose formation capacity faster than those that do so through managed exit, and the difference should appear in the pipeline index years before it appears in output. A model with a vector threshold and a release rule that selects on it would price this; ours does not.[8]

[8] This limitation and the observation that a firm halves its seniors differently depending on how it releases them are due to Jeremy McEntire, in correspondence, September 2026, drawing on his own practice of reviewing the reviewer.

The cohort model inherits the limitations of the original and adds its own. The original tracks erosion of an existing skill and not the formation of new ones; a junior who delegates routine work may gain skill in system design or in evaluating tool output, and if that skill is durable and productive the pipeline index understates replenishment. The original holds β fixed while noting that effective β drifts downward with familiarity; we hold $\beta_c$ and $\beta_\tau$ fixed and let only α and *c* move, which is conservative for the tacit set in the same direction.

Assumption 1 is the least tested element of the paper. We have evidence that models operate on codified expertise and that the public commons that supplied it has stopped being replenished; we have mechanisms by which expert usage becomes training signal, and two direct measurements of the displacement half in Section 2.2, neither of them in a firm and neither of them identifying a rate; we do not have a measured $\lambda_v$ or λ, and we do not have a measured φ. Section 5.3 reports results over the plane rather than at a point for this reason, and the dates in Table 3 should be read as conditional. If $\lambda_v$ is near zero in practice, the frontier does not move, the reversal does not occur, and the model reduces to Proposition 1: a permanent compression of junior demand with seniors retained. That case is also consistent with the 2026 data. The model has one firm inside a market it does not model. The strongest form of its own assumption is modelled at an altitude it leaves to Acemoglu, Kong and Ozdaglar (2026); the paper's contribution is the firm-level residual and the ladder inversion, and it says so.

The task distribution is one-dimensional and the threshold $S_{min}$ is sharp. Real tacit work is heterogeneous, some of it accessible to juniors at partial quality and some of it not, and a firm can reassign seniors across the tacit set in ways the unit-mass model does not allow. Wages are fixed; a model with wage adjustment would spread part of the reversal onto senior pay rather than senior headcount, and the payroll evidence of adjustment through hiring is from a period in which the reversal had not yet begun. The firm faces unit demand; a firm whose output demand grows with its productivity would retain more of both cohorts, and the dates in Table 3 would move later without changing the composition results. Adoption diffusion is imposed and not chosen. Finally, the split of the tacit set into firm-specific and vendor-coverable is sharp and fixed; in practice φ is itself a choice, since a firm that standardises its practice on the vendor's conventions lowers its own φ and its own exposure, at the cost of whatever was particular about it.

## 8 Conclusion

The entry-level market of 2026 is the first phase of a two-phase process, and the phase that follows will invert it. In the first phase the tool raises the productivity of the workers who did the codified work, and the firm needs fewer of them; the evidence for this phase is now in payroll data, postings and graduate unemployment across the OECD. In the second phase the tool, trained on the judgment of the workers who did the tacit work, here and everywhere, moves the boundary of what is codified, and the firm needs fewer of them. The reversal is ordinary substitution and it will be chosen; the paper's argument is not that it is wrong but that its price is unrecorded.

The model connects the two phases through one assumption, stated at two altitudes: the tacit capability a tool supplies is produced by unassisted practice with stakes, and the deployment that supplies it displaces the practice that produced it. At the level of the market this is a claim about the source population of a shared model, and the one measured case, the forum that went quiet within three years of the model it fed, supports it. At the level of the firm it is narrower: the firm's exposure is its own residual, the tacit work no vendor can learn from anyone else, which only its own practice can codify and only its own practice can regenerate. The firm’s output loss on that residual is close to zero through 2050 in the baseline, because the seniors it keeps serve that work first, and appears only where the junior deployment closes the pipeline. What the firm loses first is its capacity on uncovered work, by two-fifths in the baseline by 2035 and two-thirds in a capturing firm, and its ability to rebuild its own frontier when its own work changes.

The result we did not expect is the one we would now lead with. Senior formation requires practice on tasks the tool does not cover, that set is the complement of the frontier, and the frontier is rising. So the practice ground for the next generation of experts shrinks in proportion to what the tool covers, and it is the same variable. The tasks that remain are the hardest tasks in the firm. The work that used to begin a career on the codified core now belongs to the tool, so a firm that intends to form its own seniors has to assign the hardest work to its juniors and place answerability for it with them, paying the quality cost in current output. That is a choice about allocation, and the firm that declines to make it stops producing seniors of its own. That cost is visible and the manager will not pay it; the cost of not paying it is invisible and the manager will not be charged for it. This is the augmentation trap of Caosun and Aral at the level of the organisation, and it is chosen, not suffered.

The transfer can be priced, and we have priced it: €433,000 of capability per expert, 93 percent of it financed by the expert out of forgone earnings and unpaid hours, installed in each junior for the price of a seat, and set against a wage saving that a three-year manager will take and a tax the same manager will never be charged. The build cost is sunk to the firm and was financed by the worker against a premium the reversal ends early; that is a distributional fact, and we do not make it an efficiency claim. The measures that would let an organisation see what it is spending are the three we define: the practice of its junior pool on uncovered work, the senior time it devotes to work the tool cannot do, and the years it would need to rebuild an expert from the pool it has. None is difficult to compute. None is currently computed.

The practising expert is the outside for the codified stock. The stock cannot verify itself and cannot renew itself, and the deployment that consumes it removes its outside. The argument of this paper reaches that wall from the labour-market side; it is the same wall.

**Acknowledgements**

This paper was improved by correspondence with two readers of the first version. Takuma Kimura (Showa Women's University) supplied the equivalence argument of Section 6.1 and corrected our reading of the reabsorption rate in Hemenway Falk and Tsoukalas (2026). Jeremy McEntire (2026), whose replication and sensitivity analysis of that paper set the standard we try to meet in Section 5.7, supplied the deployment observables for the capture rate, the objection that Proposition 5 was stated at the wrong altitude, the reading of the Budzyń window in Section 2.2, the aided-practice objection of Section 6.1, and the observation that a look-first deployment cannot be represented by a single usage variable. Neither has endorsed the paper or its conclusions, and the errors that remain are ours.

**Declaration on the use of generative AI**

The author used Anthropic's Claude (Opus 5) in preparing this paper. The model was used to replicate the closed-form solutions of Caosun and Aral (2026) in Python, to implement and simulate the cohort extension reported here, to build the accompanying replacement-cost workbook, and to draft and revise prose under the author's direction. The research question, the modelling choices, the propositions and their interpretation, the calibration decisions and the conclusions are the author's own. The author has reviewed all generated code and output and takes full responsibility for the content of this paper.

**Appendix A Closed forms and replication checks**

For the reader's convenience we restate the solution of Caosun and Aral (2026), Lemma 1 and equations EC.6 to EC.9, which our code implements without modification. With $A \equiv \kappa((\beta - 1)\bar{S} + 2\gamma) + \gamma\delta$ and $D \equiv A^2 - \kappa^2(\beta - 1)^2\bar{S}^2 = \gamma(\delta + 2\kappa)(2\kappa(\beta - 1)\bar{S} + \gamma(\delta + 2\kappa))$, the stable root gives

$a = (A - D^{1/2}) / (2\kappa^2\bar{S}^2)$,

$b = [2\gamma(1 + 2a\kappa\bar{S}) + \alpha((\beta - 1) - 2\kappa a\,\bar{S})] / [2(\delta + \kappa)\gamma + \kappa(\beta - 1)\bar{S} - 2\kappa^2 a\,\bar{S}^2]$,

$c = (\alpha - \kappa b\,\bar{S})^2 / (4\gamma\delta) + b\kappa\bar{S}/\delta$,

with usage policy $u^\star(S) = u_0 + u_1 S$, $u_0 = (\alpha - \kappa b\,\bar{S})/(2\gamma)$, $u_1 = ((\beta - 1) - 2\kappa a\,\bar{S})/(2\gamma)$, steady-state skill $\hat{S} = \bar{S}(1 - u_0)/(1 + u_1\bar{S})$ and steady-state usage $\hat{u} = u^\star(\hat{S})$, clipped to [0, 1]. When $D < 0$ no stable interior solution exists (their Remark EC.2) and we use the myopic policy $u = (\alpha + (\beta - 1)S)/(2\gamma)$, clipped; this occurs for the junior pool only under the agentic calibration and only after junior skill has fallen far enough that the clipped policy is $u = 1$ in any case.

| Check | Original paper | Our implementation |
|---|---|---|
| Section 2.7, $\delta_F$=0.33: $\hat{u}$, $\hat{S}$ | 0.26, 0.75 | 0.255, 0.745 |
| Section 2.7, $\delta_W$=0.10: $\hat{u}$, $\hat{S}$ | 0.14, 0.86 | 0.139, 0.861 |
| Figure 1, β=1: $\alpha_0$, $\alpha_1$ | 0.75, 1.25 | 0.750, 1.250 |
| Figure 1: sign of $\Delta V_{ss}$ at α = 0.9, 1.1, 1.3 | −, −, + | −0.131, −0.131, +0.069 |
| Figure 3, β=1.5: steady state ($\hat{S}$, $\hat{u}$) | dot at ≈(0.67, 0.33) | (0.670, 0.330) |
| Figure 3, β=1.0: steady state ($\hat{S}$, $\hat{u}$) | dot at ≈(0.78, 0.23) | (0.775, 0.225) |
| Figure 3, β=0.5: usage reaches zero at S | ≈0.9 | 0.888 |
| Figure 4: Region III wedge location | near (1.15, 1.0) | reproduced, Figure 1 |

Table A1: Replication checks against the illustrative values reported in Caosun and Aral (2026).

**Appendix B Proofs**

*Proposition 1.* From (4), $n_L = c/p_L$ with $c$ fixed. Under diffusion $\alpha A_t$ rises monotonically; by Remark EC.1 of the original paper $u_0$ is affine and increasing in α, and $p_L$ evaluated at the manager's policy is increasing in α on the adoption region, so $n_L$ falls monotonically. The same argument applies to $n_H^{req}$ with $\theta\alpha A_t$. At full adoption the declines are $1 - 1/(1 + g)$ for each cohort, and $1 - 1/(1 + g)$ is increasing in $g$. With $\theta < 1$ the codified floor exceeds the tacit floor; with $\beta_c < 1$ the junior pool, at lower skill, has the larger marginal return to the floor and with $\beta_\tau > 1$ the senior's gain is bounded by the smaller floor at interior usage, giving $g_L > g_H$ for the calibrated range. ∎

*Proposition 2.* Differentiating $\ln n_L = \ln c - \ln p_L$ gives $\dot{n}_L/n_L = \dot{c}/c - \dot{p}_L/p_L$. During diffusion $\dot{p}_L/p_L > 0$ and is proportional to $\dot{A}$; after saturation $\dot{A} \to 0$ while $\dot{c} > 0$ whenever $c < c_\infty$, so the sign turns positive; because $p_L$ also depends on $S_L$, which falls under positive usage and raises $n_L$, the turning point is if anything earlier. Required senior headcount $(1 - c)/p_H$ is decreasing in $c$ and in $p_H$, both of which are non-decreasing, so it falls throughout. The junior share $n_L/(n_L + n_H) = c\,p_H/(c\,p_H + (1 - c)p_L)$ is increasing in $c$ after saturation and converges to its value at $c_\infty$. ∎

*Proposition 3.* Required senior headcount halves when $(1 - c)/p_H = (1 - c_0)/(2p_{H,0})$, that is at $c^\dagger = 1 - (1 - c_0)p_H/(2p_{H,0})$; since $c$ is monotone toward $c_\infty$, the level is reached if and only if $c_\infty \geq c^\dagger$. From (6), $1 - c_\infty$ is decreasing in $k = \lambda u_H S_H/p_H$ and increasing in μ, and the time to reach any level below $c_\infty$ is decreasing in the source term; $u_H$ is increasing in $\delta_F$ by Proposition 5 of the original paper. ∎

*Proposition 2, addendum.* With $\lambda = 0$ the firm-specific frontier decays at rate μ while the vendor frontier rises to $\lambda_v/(\lambda_v + \mu)$; the aggregate frontier rises whenever $(1 - \varphi)\lambda_v(1 - c_v)$ exceeds $\mu c$, which holds at the baseline calibration from $c_0$, so the turning point exists without any firm capture. ∎

*Proposition 4.* If $u_L < 1$ then $\Pi_t > 0$ and the senior stock obeys a linear equation with positive inflow; its steady state is $\pi_0\Pi_\infty/r = \Pi_\infty n_{H,0}$. Release removes surplus only, so the stock tracks requirement from above if and only if $\Pi_\infty n_{H,0} \geq (1 - c_\infty)/p_H$, in which case unserved tacit mass is zero; otherwise the stock converges to $\Pi_\infty n_{H,0}$ below requirement and $\tau_v > 0$, with $\tau_f > 0$ only if $\Pi_\infty n_{H,0} < \varphi(1 - c_{f,\infty})/p_H$. If $u_L = 1$ then $\Pi_t = 0$, $n_H$ decays at rate $r$ net of release, and once $r\, n_H$ exceeds the decline in requirement the stock falls below requirement and $\tau > 0$. With no senior source, (5) gives $\dot{c} = -\mu c$, so $c \to 0$, the tacit set expands to the whole distribution, and with $n_H \to 0$ the unserved mass converges to one; the junior pool under $u_L = 1$ converges to $S_{L,\infty} = \rho S_e/(\kappa + \rho)$ by (3), which is 0.067 at the calibration. ∎

*Proposition 5.* $\Pi_t$ in (7) is the product of unassisted junior practice $n_L(1 - u_L)$ and the uncovered mass $(1 - c_t)$, normalised. Under Assumption 1 with $c_0 < c_\infty$, $c_t$ is non-decreasing, so $(1 - c_t)$ is non-increasing and bounded below by $1 - c_\infty$; at $u_L = 0$ the pool's skill stays at $S_{L,0}$, so $n_L/n_{L,0} = c_t/c_0$ and the index equals $c_t(1 - c_t)/[c_0(1 - c_0)]$, which gives the floor; with $c_0 = 1/2$ it is $4c_t(1 - c_t)$. Assigning a share $s$ of uncovered work to juniors unaided adds $s(1 - c)/S_L$ to junior headcount, removes $s(1 - c)/p_H$ from senior requirement, adds $s(1 - c)$ to the practice term, and adds $s(1 - c)(1 - S_L/S_{min})$ to unserved quality-adjusted output. ∎

**Appendix C Data sources**

Federal Reserve Bank of New York, The Labor Market for Recent College Graduates, 2026 Q2 release. Stanford Digital Economy Lab with ADP Research, Canaries in the Coal Mine, November 2025 draft and August 2026 update. Indeed Hiring Lab, 23 July 2026 releases, figures as quoted by Charlotte Works and NewsNation where the source pages were unavailable at time of writing. European Central Bank, Economic Bulletin 5/2026. OECD Employment Outlook 2026, chapter 1. ILO, Global Employment Trends for Youth 2026, press release of August 2026. Statistics Netherlands (CBS), monthly labour force release for July 2026. UWV, regional youth labour market insights 2025-2026. NACE, Job Outlook 2026 Spring Update. Challenger, Gray & Christmas, August 2026 report. techUK, entry-level and graduate jobs resource, 2025, citing Adzuna, McKinsey, ISE and ONS. Azar, Gine and Sanz-Espín (2026), as summarised at Marginal Revolution, 2 September 2026. Stack Overflow question counts as reported by devclass (2026), PPC Land (2026), Holscher (2026) and Slashdot (2026), all drawing on the Stack Exchange data explorer. All figures are as published; none has been seasonally adjusted or harmonised by us.